%% file: main.tex
\documentclass[aps,prl,10pt,twocolumn,superscriptaddress,nofootinbib,floatfix,raggedbottom]{revtex4-2}
\usepackage{amsmath,amssymb,bm,mathtools,booktabs}
\usepackage{tikz}
\usetikzlibrary{arrows.meta,calc,positioning}
\definecolor{resblue}{HTML}{27678A}
\definecolor{resrust}{HTML}{BC593D}
\definecolor{resgray}{HTML}{66717A}
\usepackage[T1]{fontenc}
\usepackage{CJKutf8}
\usepackage{lmodern}
\usepackage{microtype}
\usepackage{hyperref}
\hypersetup{pdftitle={Analytic Dual Resonance and the String Bootstrap},pdfauthor={Q. Cao}}
\newcommand{\A}{\mathcal A}
\newcommand{\dd}{\mathrm d}
\newcommand{\Res}{\operatorname*{Res}}
\newcommand{\bmm}{\boldsymbol m}
\newcounter{smapp}

\newcommand{\appref}[1]{\hyperref[#1]{App.~\ref*{#1}}}
\newcommand{\apprange}[2]{Apps.~\ref{#1}--\ref{#2}}
\begin{document}
\clubpenalty=1000
\widowpenalty=1000
\displaywidowpenalty=1000

\begin{CJK*}{UTF8}{gbsn}
\title{Analytic Dual Resonance and the String Bootstrap}
	\author{Qu Cao(曹趣)}
		\email{caoqu@westlake.edu.cn}
		\affiliation{Department of Physics, School of Science, Westlake University, Hangzhou 310030, China}

\begin{abstract}
Dual resonance reconstructs a scattering amplitude from pole expansions
in different channels. We propose analytic dual resonance (ADR),
requiring that the complete leading pole sum at an integer resonance
admit an absolutely convergent local contour integral. ADR yields
one-dimensional pairing relations between residues in arbitrary
triangulations. Together with an integer planar spectrum, a joint
crossing-spin bound, and complex Regge decay permitting unsubtracted
reconstruction, these relations uniquely fix the scalar string amplitude
for every \(n\ge5\). ADR also predicts higher-dimensional pinch
relations, which we examine at five and six points. These results connect
the analytic structure of dual resonance to the uniqueness of string
interactions.
\end{abstract}
\maketitle
	\end{CJK*}
\paragraph{Introduction.---}
In a Feynman-diagram expansion, exchanges in different channels are
added. In a dual-resonant amplitude, the complete exchange sum in one
channel already contains the poles of the crossed channels. This is the idea behind
resonance--Regge duality~\cite{DolenHornSchmid}, realized by the
Veneziano amplitude~\cite{Veneziano} and its multiparticle
extensions~\cite{KobaNielsen}. What fixes the interactions of this
infinite tower?

String bootstraps address this question through high-energy and
Regge-trajectory constraints~\cite{MatsudaUniqueness,AsymptoticUniqueness,StringyMaximalSpin}, residue
zeros~\cite{StringBootstrap,StringsAlmostNothing}, and multiparticle
factorization~\cite{Rigidity,HigherPointConstraints}. Positivity combined
with monodromy or low-lying spectral data constrains four-point couplings
near string values~\cite{CarvingOpenString,StringEFTBootstrap,FlatteningEFTHedron,StringIslands}.
Analytic four-point uniqueness results are also known~\cite{AnalyticVeneziano}.
Complementary approaches use splitting~\cite{SplittingIslands}, EFT
consistency~\cite{MaximalSUSYString}, unitarity~\cite{TreeUnitarity},
multipositivity~\cite{Multipositivity}, and gauge-invariant
operators~\cite{BosonicEffectiveUniqueness}.

At five points, Regge zeros and cyclicity fix a polynomial
residue ansatz~\cite{StringsAlmostNothing}. Residue matching and
monodromy also provide a route to arbitrary-point tachyon
amplitudes~\cite{StringTargetSpace}.
Dual resonance alone, however, admits
deformations~\cite{StringyDynamics,VenezianoVariations}.
Four-point residue matching~\cite{StringyDynamics} suggests where to look
for a stronger condition: the channel expansions should remain compatible
when their poles meet. Equality of the full sums alone does not ensure
this local compatibility.

In this Letter, we propose \emph{analytic dual resonance} (ADR): the sum
of leading pole terms at a collision must admit an absolutely convergent
local contour integral. At a two-pole collision, this condition equates
crossed-channel residues. From ADR we derive general
one-dimensional \emph{pairing} conditions for triangulation flips at
arbitrary multiplicity.
Together with an integer planar spectrum, a joint
crossing-spin bound, and Regge decay permitting unsubtracted
reconstruction, these conditions fix the scalar string amplitude for
every \(n\ge5\). The proof uses the full degree-bounded polynomial
space, without prescribed zeros, monodromy relations, or lower-point
factorization data. ADR also gives higher-dimensional pinch relations,
which we test alongside the five- and six-point bootstrap examples.

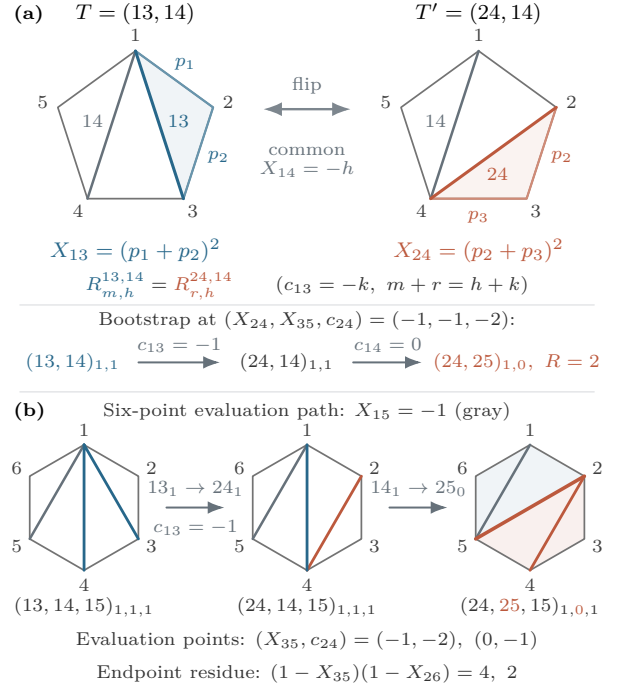
\begin{figure}[t]
\centering
\input{figures/channel_flip}
\caption{Paired flips.
(a) Momenta label polygon sides; \(13\leftrightarrow24\) preserves channel \(14\).
Levels \(m,r,h,k\ge0\) are integers; other kinematics are generic.
(b) With \(X_{24}=X_{46}=-1\) and \(c_{25}=c_{35}=0\), the second flip
uses \((c_{26},c_{36})=(0,-1),(-1,0)\).
Arrow subscripts denote levels.}
\label{fig:channels}
\end{figure}

\paragraph{Kinematics and dual resonance.---}
We bootstrap color-ordered tree amplitudes \(\A_n\) of massless scalars. Define
\begin{equation}
 X_{ij}=(p_i+p_{i+1}+\cdots+p_{j-1})^2,\quad c_{ij}=-2p_i\cdot p_j.
 \label{eq:kinematics}
\end{equation}
The planar Mandelstam variable \(X_{ij}\) labels the diagonal \((ij)\)
of a polygon whose side \((i,i+1)\) carries \(p_i\)
[Fig.~\ref{fig:channels}]. A triangulation is dual to a planar cubic
tree, with diagonals crossing internal propagators.
The ABHY relations~\cite{ABHY} read
\begin{equation}
 c_{ij}=X_{ij}+X_{i+1,j+1}-X_{i,j+1}-X_{i+1,j},
 \label{eq:cdef}
\end{equation}
with cyclic indices and \(X_{i,i+1}=p_i^2=0\).
We treat the planar invariants as independent; fixing suitable \(c_{ij}\)
defines an affine slice. We take an integer planar spectrum as input:
the only finite singularities are simple channel poles at
\(X_{ij}=-m\), \(m=0,1,\ldots\).
Nonzero multiple residues require noncrossing channels.
The interactions of this tower are to be determined.

A triangulation \(T\) selects \(n{-}3\) compatible channels. Let
\(R^T_{\boldsymbol m}\) denote their maximal residue at
\(X_e=-m_e\), with subscripts in channel order.\footnote{We normalize one all-zero-level maximal residue to unity at each multiplicity.}
An ordered pole expansion takes the schematic form
\begin{equation}
 \A_n=\sum_{\boldsymbol m\ge0}
 \frac{R^T_{\boldsymbol m}}{\prod_{e\in T}(X_e+m_e)}+B_n^T.
 \label{eq:pole-reconstruction}
\end{equation}
The sum is analytically continued from its convergence domain, with
channels outside \(T\) fixed. The boundary term \(B_n^T\) can include
terms with fewer poles in \(T\); complex Regge decay removes it in our
fan reconstruction (\appref{sm:V}). Ordinary dual resonance states that,
once boundary terms vanish, these complete sums describe the same
amplitude for every choice of \(T\).

At four points, \(c_{13}=X_{13}+X_{24}\), and \(R_m(c_{13})\)
denotes the residue at \(X_{13}=-m\). For a cyclic amplitude,
dual resonance reads
\begin{equation}
 \A_4=\sum_{m\ge0}\frac{R_m(X_{24}-m)}{X_{13}+m}
     =\sum_{m\ge0}\frac{R_m(X_{13}-m)}{X_{24}+m}.
 \label{eq:DR4}
\end{equation}
Equality of
the sums does not by itself determine how individual residues behave
when poles collide. Can a common analytic condition relate the individual residues across levels and triangulations?

\paragraph{A four-point clue.---}
Our clue comes from the scalar string amplitude \(\A_n^{\rm str}\), the
ordered Koba--Nielsen disk integral with the matching Parke--Taylor
factor~\cite{Rigidity,OpenStringModern,ZDiskIntegrals,AbelianZTheory,NonAbelianZTheory,StringyCanonicalForms}.
At four points, the Veneziano amplitude and its residues are
\begin{equation}
 \begin{aligned}
 \A_4^{\rm str}
 =\frac{\Gamma(X_{13})\Gamma(X_{24})}{\Gamma(c_{13})},\, R_m(c_{13})=\frac{(-1)^m(c_{13})_m}{m!},
 \end{aligned}
 \label{eq:beta}
\end{equation}
where \((a)_m=a(a+1)\cdots(a+m-1)\), with \((a)_0=1\).
At \(c_{13}=-k\), with \(k\ge0\) integer, the residues obey
\begin{equation}
 \begin{aligned}
 R_a(-k)&=R_{k-a}(-k),&&0\le a\le k,\\
 R_a(-k)&=0,&&a>k.
 \end{aligned}
 \label{eq:pair-zero}
\end{equation}
The second relation gives the truncation zeros~\cite{StringBootstrap}:
on \(X_{13}=-a\), they lie at \(X_{24}=1,\ldots,a\)~\cite{StringsAlmostNothing}.
Together with the first relation and the degree bound, they fix the
four-point residues and hence the unsubtracted amplitude (\appref{sm:I}).
We focus on the first relation, which we call
\emph{pairing}~\cite{StringyDynamics}: crossed-channel residues agree
where their poles meet.

To see how pairing cancels a pole, restrict to \(c_{13}=-k\) and write
\(X_{13}=-m+u\), \(X_{24}=-r-u\), with \(m+r=k\).
The absence of an incompatible double residue gives
\begin{equation}
 \Res_{u=0}\A_4(-m+u,-r-u)=R_m(-k)-R_r(-k).
 \label{eq:local-poles}
\end{equation}
The Veneziano amplitude vanishes along this slice because
\(1/\Gamma(c_{13})=0\)~\cite{HiddenZeros}; pairing removes the apparent
pole at the intersection. For a general dual-resonant amplitude, taking residues term by term need not commute with continuing the infinite sum to this collision~\cite{StringyDynamics}.\footnote{See \appref{sm:I} for a satellite counterexample.}

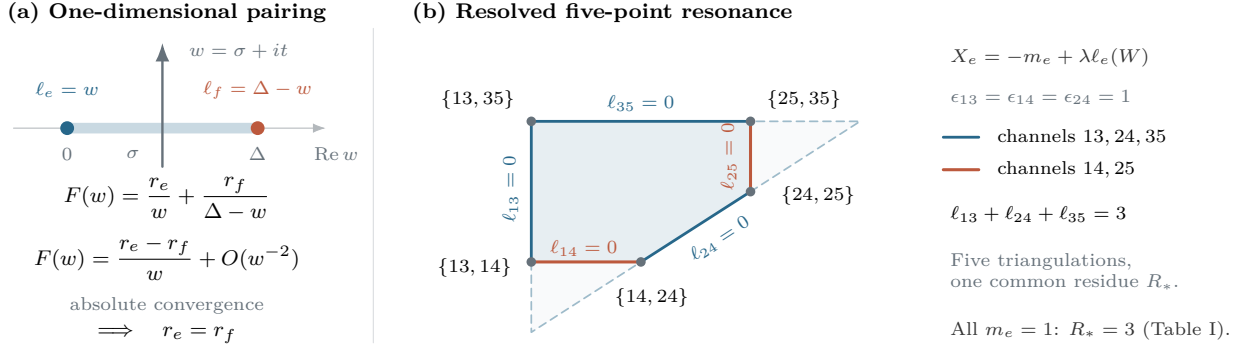
\begin{figure*}[t]
\centering
\input{figures/pinch_geometry}
\caption{Local resonance geometry. (a) The contour \(w=\sigma+it\),
\(0<\sigma<\Delta\), runs upward between the two poles. Equal
residues cancel the \(1/w\) tail. (b) The resolved five-point collision
in displacement space, with \(\epsilon_{13}=\epsilon_{14}=\epsilon_{24}=1\),
independent of the integer levels:
\(\ell_{13}=W_1\), \(\ell_{14}=W_2\),
\(\ell_{24}=1-W_1+W_2\), \(\ell_{25}=2-W_1\),
\(\ell_{35}=2-W_2\).
The three blue pole lines bound a triangle, truncated by the two
remaining pole lines (rust) to a pentagon. Its vertices label compatible double
residues at the same integer resonance, not successive flips.}
\label{fig:pinch}
\end{figure*}

\paragraph{Analytic dual resonance.---}
Pairing has a simple local analytic interpretation.
Resolve the four-point collision by writing
\(X_e=-m+\lambda w\), \(X_f=-r+\lambda(\Delta-w)\), with
\(\Delta>0\). The coefficient of \(\lambda^{-1}\) is the leading pole sum
\begin{equation}
 F(w)=\frac{r_e}{w}+\frac{r_f}{\Delta-w}
     =\frac{r_e-r_f}{w}+O(w^{-2}).
 \label{eq:local-period-tail}
\end{equation}
Here \(r_e,r_f\) are the collision residues. On the vertical contour
\(w=\sigma+it\), \(0<\sigma<\Delta\) [Fig.~\ref{fig:pinch}(a)],
absolute convergence of the contour integral is equivalent to \(r_e=r_f\):
the \(1/w\) tails cancel, leaving \(1/w^2\) decay.

For a collision with \(d\) independent kinematic shifts \(W\), write
\(X_e=-m_e+\lambda\ell_e(W)\), with the \(\ell_e\) obeying
the ABHY relations. At fixed \(W\), define the
complete leading pole sum \(F\) by
\begin{equation}
 \mathcal B=\lambda^{-d}F(W)+O(\lambda^{-d+1}).
 \label{eq:resonance-leading}
\end{equation}
Here \(\mathcal B\) is the amplitude or its residue in common channels,
taken before the collision at generic kinematics. Thus \(d=1\) for a
two-pole collision and \(d=n-3\) when all planar channels reach poles.
Choose real \(\sigma\) with all
\(\ell_e(\sigma)>0\), so the contour avoids the poles. Our
\emph{ADR hypothesis} is
\begin{equation}
\int_{\mathbb R^d}|F(\sigma+it)|\,\dd^dt<\infty.
 \label{eq:adr}
\end{equation}
The resonance limit precedes integration. This local condition leaves
regular terms free; it is not a high-energy limit of the full amplitude.
We impose it for one admissible positive resolution and contour at each
two-pole flip locus and complete integer resonance specified in
\apprange{sm:I}{sm:II}. The resulting residue relations ensure convergence
for every such choice.

\paragraph{Pairing at higher points.---}
The same cancellation applies to every triangulation flip.
Exchange \(e\) for its crossing partner \(f\),
preserving common channels \(S\):
\(T=S\cup\{e\}\), \(T'=S\cup\{f\}\).
Taking residues in the common channels at levels \(\boldsymbol h\), with other
kinematics generic, leaves a two-pole problem for \(\mathcal B(X_e,X_f)\).

For example, Fig.~\ref{fig:channels}(a) shows a flip with the common channel
at \(X_{14}=-h\). Levels \(m,r\) in channels \(13,24\) meet on
\(c_{13}=-k\) when \(m+r=h+k\). The other locus for this flip sets
\(c_{24}=-k_1,c_{25}=-k_2\), with \(m+r=k_1+k_2\).

In general, the flip defines four consecutive particle groups
\(I,J,K,L\), labeled in the diagram below. For \(Q=I\times K\), set each
\(c_{ij}=-k_{ij}\), \(k_{ij}\in\mathbb Z_{\ge0}\).
Summing Eq.~\eqref{eq:cdef} gives
\(X_e+X_f=-h_Q-\sum_Q k_{ij}\), where
\(h_Q=m_J+m_L\). Here \(m_A\) is the level of the common channel enclosing
group \(A\), or zero for a single particle. The other choice is
\(Q=J\times L\), with \(h_Q=m_I+m_K\) (\appref{sm:I}).
For \(m,r\ge0\), ADR cancels the \(1/u\) pole in
\(\mathcal B(-m+u,-r-u)\), giving
\begin{equation}
 \text{\input{figures/pairing_equation}\unskip}\quad
 m+r=h_Q+\sum_Qk_{ij}.
 \label{eq:resolved-pair}
\end{equation}
Dashed diagonals cross solid dual propagators; the attached
currents are unchanged.
The exchanged level is listed first in each residue. The relation holds for both choices
of opposite groups, every flip, and all common-channel levels.
Each \(c_{ij}\) is fixed individually; other independent variables remain unrestricted.

\paragraph{Five- and six-point bootstrap.---}
To turn pairing into a bootstrap, we need a bound on the residue
polynomials. We impose the joint \emph{crossing-spin bound}
\begin{equation}
 \deg_e R^T_{\boldsymbol m}\le m_e.
 \label{eq:bootstrap}
\end{equation}
Here \(\deg_e\) counts total degree in planar variables whose diagonals
cross \(e\). A spin-\(J\) exchange joins two currents
through at most \(J\) momentum contractions across that channel.
Related bounds appear in Ref.~\cite{StringTargetSpace}.
The bound is motivated, but not implied, by \(J\le m_e\).
At level zero it removes dependence on all crossing channels; it does
not prescribe a factorized residue.

We begin at five points. Write \(R_{a,b}\equiv R^{13,14}_{a,b}\), with
\(a,b\) the respective levels in channels \(13,14\).
For \(b=0\), the degree bound leaves a polynomial only in \(X_{24}\).
Pairing relates its values at nonpositive integers to the corresponding
boundary polynomials in the other triangulations. These boundaries must
also be restrictions of the same level-\((a,1)\) residues. Their
compatibility fixes the boundary polynomials at every level up to two constants
(\appref{sm:III}):
\begin{equation}
 R^{13,14}_{a,0}=t_5\frac{(1-X_{24})_a}{a!}+q_5.
 \label{eq:solved-axis}
\end{equation}
The constants \(t_5,q_5\) are common to all levels and triangulations,
even without imposing cyclic symmetry.
Regge decay and normalization give
\(t_5=1,q_5=0\) (\appref{sm:V}); hence
\(R^{13,14}_{1,0}=1-X_{24}\).

With the zero-level data fixed, now consider positive levels. In coordinates
\((X_{24},X_{35},c_{24})\), the bounds are
\(\deg_{X_{24},c_{24}}R_{a,b}\le a\) and
\(\deg_{X_{35},c_{24}}R_{a,b}\le b\).
At the lowest positive levels, \((a,b)=(1,1)\), there are five
undetermined coefficients:
\begin{equation}
R^{13,14}_{1,1}=\alpha_0{+}\alpha_1X_{24}{+}\alpha_2X_{35}{+}\alpha_3X_{24}X_{35}{+}\alpha_4c_{24}.
 \label{eq:ansatz11}
\end{equation}
At these levels, \(c_{13}=X_{24}\) and \(c_{35}=X_{35}\).
The single-cell \(k=0\) pairings with \(R^{24,14}_{0,1}\) and
\(R^{13,35}_{1,0}\) therefore fix the full boundaries \(X_{24}=0\)
and \(X_{35}=0\), with other variables generic.
Both partner residues are \(1-X_{25}\); using
\(X_{25}=X_{24}+X_{35}-c_{24}\), these become
\(1-X_{35}+c_{24}\) and \(1-X_{24}+c_{24}\), respectively.
Only \(\alpha_3\) is undetermined. At
\((X_{24},X_{35},c_{24})=(-1,-1,-2)\), the two flips in
Fig.~\ref{fig:channels}(a) give
\(R^{24,25}_{1,0}=1-X_{35}=2\).
Thus \(1+\alpha_3=2\), and
\begin{equation}
 R^{13,14}_{1,1}=(1-X_{24})(1-X_{35})+c_{24}.
 \label{eq:R11}
\end{equation}
The induction below fixes the result at all levels:
\begin{equation}
 R^{13,14}_{a,b}=\sum_{r=0}^{\min(a,b)}
 \frac{(1-X_{24})_{a-r}(1-X_{35})_{b-r}(c_{24})_r}
 {(a-r)!(b-r)!r!}.
 \label{eq:five-all}
\end{equation}
The \(r=0\) term factorizes into four-point residues;
terms with \(r>0\) couple the channels through \(c_{24}\).

At six points, the same bound allows 13 coefficients for the triple residue
\(R=R^{13,14,15}_{1,1,1}\).
Using the level-zero results (\appref{sm:IV}), the flips
\(13_1\to24_0\), \(15_1\to46_0\) give
\begin{equation}
 \begin{aligned}
 R|_{X_{24}=0}&=(1-X_{46})(1-X_{35}+c_{24})+c_{35}+c_{25},\\
 R|_{X_{46}=0}&=(1-X_{24})(1-X_{35}+c_{35})+c_{24}+c_{25}.
 \end{aligned}
 \label{eq:six-boundaries}
\end{equation}
The remaining ambiguity vanishes on both boundaries, hence is divisible
by \(X_{24}X_{46}\); the degree bound restricts the quotient to
\(\alpha+\beta X_{35}\), giving
\begin{equation}
 \begin{split}
 R={}&(1-X_{35})(1-X_{24}-X_{46})\\
 &+c_{24}(1-X_{46})+c_{35}(1-X_{24})+c_{25}\\
 &+X_{24}X_{46}(\alpha+\beta X_{35}).
 \end{split}
 \label{eq:six-reduced}
\end{equation}
To fix \(\alpha,\beta\), follow the flips in
Fig.~\ref{fig:channels}(b) to \(R^{24,25,15}_{1,0,1}\).
Set \(X_{24}=X_{46}=-1\), \(c_{35}=c_{25}=0\), and take
\((X_{35},c_{24})=(-1,-2),(0,-1)\).
The level-zero channel \(25\) separates two quadrilaterals.
Pairing and the joint degree bound fix this endpoint residue to
\((1-X_{35})(1-X_{26})\), as derived in \appref{sm:IV},
without assuming factorization.
The ABHY relations give \(X_{26}=-1\) at both points, hence the values
\(4\) and \(2\).
Equation~\eqref{eq:six-reduced} then gives
\(2+\alpha-\beta=4\) and \(1+\alpha=2\), so
\(\alpha=1,\beta=-1\) and
\begin{equation}
 \begin{split}
 R^{13,14,15}_{1,1,1}={}&(1-X_{24})(1-X_{35})(1-X_{46})\\
 &+c_{24}(1-X_{46})+c_{35}(1-X_{24})+c_{25}.
 \end{split}
 \label{eq:six-unit}
\end{equation}
Thus pairing determines both the product term and the terms coupling
different channels, reproducing the string residue.

\paragraph{Uniqueness at arbitrary multiplicity.---}
For every \(n\ge5\), the integer planar spectrum, ADR pairing conditions, joint
crossing-spin bound, and Regge condition imply
\(\A_n=\A_n^{\rm str}\). The proof first classifies the maximal
residues. Pairing on both loci of every flip and the joint degree bound give
\begin{equation}
 R^T_{\boldsymbol m}
 =t_n R^{T,\rm str}_{\boldsymbol m}+q_n,\qquad n\ge5.
 \label{eq:classification}
\end{equation}
The constants \(t_n,q_n\) are common to all levels and triangulations.
The constant \(q_n\) shifts every residue; it is not a contact interaction.

We induct on multiplicity and total level, with five and six points
treated in \apprange{sm:III}{sm:IV}. At a zero-level cut, the degree bound removes
crossing variables, leaving lower-point pairing on each side.
Cuts removing a boundary triangle
inherit the \((n-1)\)-point classification. Comparing compatible cuts
fixes common constants \(t_n,q_n\) and then determines the residues on
the remaining zero-level cuts. The six-point split into two quadrilaterals requires
the separate argument illustrated above.

For positive levels, subtract the right side of
Eq.~\eqref{eq:classification} and induct on \(M=\sum_{e\in T}m_e\).
Replacing a flipped level \(m\) by \(r<m\) lowers \(M\), so the
corresponding difference in the partner triangulation vanishes by induction.
For a flip with \(h_Q<m\), these paired values force
polynomial factors that exhaust the allowed crossing degree.
The quotient therefore cannot depend on
variables crossing the exchanged channel. Two same-level comparisons in
a five-point subdiagram then impose incompatible restrictions unless
this quotient vanishes, even at arbitrary attachment levels. Vanishing
propagates across same-level flips whose exchanged
level is at least the level sum of one pair of opposite sides.
\appref{sm:III} constructs such paths from every triangulation.

Finally, we use Regge decay in the form that permits unsubtracted
pole reconstruction~\cite{StringyDynamics,StringBootstrap,OpenStringModern}.
Its precise statement is Eq.~\eqref{app:prefix-regge}: a mean absolute
bound on complex circles in an ordered fan, also after fixed-level
residues. It excludes the constant tower \(q_n\) and the boundary terms
in Eq.~\eqref{eq:pole-reconstruction}; normalization fixes \(t_n=1\).
Successive Cauchy reconstruction then gives \(\A_n=\A_n^{\rm str}\),
as detailed in \appref{sm:V}. \appref{sm:VI} verifies the assumptions
for the string solution.

\paragraph{Higher-dimensional resonances.---}
The bootstrap used only two-pole collisions. ADR also constrains simultaneous collisions in several kinematic directions. At the complete integer resonances specified below, it requires the maximal residues of all triangulations to agree. The leading pole sum then has the form of a
planar \(\phi^3\) tree amplitude. To locate these collisions, use the ABHY identity 
\begin{equation}
 \sum\nolimits_{i=1}^{n-2}X_{i,i+2}
 =\sum\nolimits_{\substack{1\le i<j\le n-1\\j-i\ge2}}c_{ij}.
 \label{eq:short-circuit}
\end{equation}
The sum runs over \(n-2\) consecutive two-particle channels and is fixed
under the \(n-3\) independent fan shifts, allowing a simplex pinch.
Other channels that reach physical poles must be included in the same
collision.

Set \(c_{ij}=-k_{ij}\), with \(k_{ij}\ge0\) integer, and place
these two-particle channels at physical poles, with
\(\sum_i m_{i,i+2}=\sum_{ij}k_{ij}\). Require every derived planar
level to be nonnegative. At five points,
\begin{equation}
 \begin{aligned}
 &m_{13}+m_{24}+m_{35}=k_{13}+k_{14}+k_{24},\\
 &m_{14}=m_{13}+m_{24}-k_{13},\,\,
 m_{25}=m_{24}+m_{35}-k_{24}.
 \end{aligned}
 \label{eq:five-physical-node}
\end{equation}
Thus all five channels reach poles.
The three-dimensional six-point collision obeys
\begin{equation}
 \begin{split}
 m_{13}{+}m_{24}{+}m_{35}{+}m_{46}=k_{13}{+}k_{14}{+}k_{15}{+}k_{24}{+}k_{25}{+}k_{35}.
 \end{split}
 \label{eq:six-physical-node}
\end{equation}
The ABHY relations fix all nine planar levels.

Resolve the collision by \(X_e=-m_e+\lambda\ell_e\) and
\(c_{ij}=-k_{ij}+\lambda\epsilon_{ij}\), with \(\epsilon_{ij}>0\).
In fan coordinates \(W_i=\ell_{1,i+2}\), the region \(\ell_e(W)>0\)
is the \((n-3)\)-dimensional associahedron~\cite{ABHY}.
At five points, its vertices label compatible double poles
[Fig.~\ref{fig:pinch}(b)].

In the leading pole sum, triangulation \(T\) contributes
\(r_T/\prod_{e\in T}\ell_e\), with \(r_T\) its maximal residue at resonance.
Schwinger parameters identify the inverse Fourier transform of
\(F(\sigma+it)\): its limit at the origin within the inward normal
cone of vertex \(T\) is \(r_T\). ADR makes this transform continuous,
so all such limits agree (\appref{sm:II}):
\begin{equation}
 F(W)=R_*\sum_T\frac1{\prod_{e\in T}\ell_e(W)}, \quad r_T=R_*.
 \label{eq:physical-overlap}
\end{equation}
The sum is the unit-coupling planar \(\phi^3\) tree amplitude in
\(\ell_e\). All mass-level dependence lies in \(R_*\), which equals
the local contour integral of \(F\) with measure
\(\dd^{n-3}W/(2\pi i)^{n-3}\) (\appref{sm:II}).

With all planar levels set to one, Eq.~\eqref{eq:R11} gives \(R_*=3\)
at five points; at six points, \(c_{24}=c_{35}=-1,c_{25}=0\),
so Eq.~\eqref{eq:six-unit} yields \(R_*=4\).
Table~\ref{tab:higher-resonance} summarizes checks across all
triangulations, including higher levels, using independently
computed string residues.

The five-point hypergeometric representation~\cite{OpenStringModern}
independently verifies Eq.~\eqref{eq:physical-overlap} at all allowed levels:
its analytically continued tail cancels the apparent incompatible double
pole in \(24,35\) and supplies the \(25\) terms (\appref{sm:II}).
At these complete integer resonances, paired flips already equate all
maximal residues and ensure absolute convergence. The higher-dimensional
conditions thus provide consistency checks, not additional bootstrap inputs.

\begin{table}[!htbp]
\caption{Complete resonances of dimension \(d=n-3\), with all planar
channels at \(X_e=-m\) and all triangulations sharing residue \(R_*\).}
\label{tab:higher-resonance}
\centering
\setlength{\tabcolsep}{3pt}
\renewcommand{\arraystretch}{1.15}
\begin{tabular}{@{}cccccc@{}}
\toprule
\(n\)&\(d\)&\(m\)&\(k_{ij}\)&Triangulations&\(R_*\)\\
\midrule
5&2&1&\((1,1,1)\)&5&3\\
5&2&2&\((2,2,2)\)&5&19\\
6&3&1&\((1,0,1,1,0,1)\)&14&4\\
6&3&2&\((2,0,2,2,0,2)\)&14&48\\
\bottomrule
\end{tabular}
\par\smallskip\footnotesize
Five points: \((k_{13},k_{14},k_{24})\).\\
Six points: \((k_{13},k_{14},k_{15},k_{24},k_{25},k_{35})\).
\end{table}

\paragraph{Conclusion and Outlook.---}
Dual resonance requires different channel expansions to describe the same
amplitude. ADR extends this compatibility to colliding poles through a
local convergence condition. Its one-dimensional pairing relations,
together with an integer planar spectrum, a joint crossing-spin bound,
and Regge decay, uniquely fix the scalar string amplitude for every
\(n\ge5\). The five- and six-point examples show how pairing determines
the interactions between channels; the proof applies at arbitrary
multiplicity.\footnote{These assumptions
do not fix an isolated four-point amplitude. The Veneziano amplitude is nevertheless
recovered as the level-zero residue in \(X_{14}\) of the five-point solution.}

The higher-dimensional predictions of ADR give a further test of this
picture. At complete integer resonances, all triangulations share one
residue, multiplying a planar \(\phi^3\) pole sum. The five- and six-point
string amplitudes satisfy these relations, including resonances where
mutually incompatible channels reach their poles together.

What physical mechanism enforces ADR? A global contour
representation may connect pairing and higher-dimensional pinches,
with Mellin analysis~\cite{Yuan} and Lorentzian integration
cycles~\cite{LorentzianContours} providing possible starting points.
Extensions to closed strings~\cite{ClosedStringBootstrap} and amplitudes
with spinning external states would test how far the same analytic
condition constrains string interactions.

\paragraph{Acknowledgements.---}
	We thank Yichao Tang and Ellis Ye Yuan for stimulating discussions. The work of Q.C. is supported by the Westlake Fellows Program at Westlake University. OpenAI's GPT-6 model, accessed through Codex, assisted with derivations, computational checks, and manuscript preparation; all AI-assisted material and scientific conclusions were reviewed and verified by the author, who takes full responsibility for the content.

\bibliographystyle{apsrev4-2}
\bibliography{dual_resonance_references}
\onecolumngrid
\appendix
\input{dual_resonance_pinch_supplement}
\end{document}

%% file: figures/pairing_equation.tex
\begin{tikzpicture}[x=1cm,y=1cm,baseline=(leftres.base),
 line cap=round,line join=round,>=Latex]
 \path[use as bounding box] (-.82,-.42) rectangle (3.32,1.57);
 \foreach \x in {0,2.5} {
  \draw[resgray!50,line width=.4pt]
   (\x,1.45)--(\x+.75,.80)--(\x,.15)--(\x-.75,.80)--cycle;
  \node[text=resgray!90!black,font=\footnotesize,inner sep=0pt]
   at (\x-.64,1.43) {$I$};
  \node[text=resgray!90!black,font=\footnotesize,inner sep=0pt]
   at (\x-.64,.17) {$J$};
  \node[text=resgray!90!black,font=\footnotesize,inner sep=0pt]
   at (\x+.64,.17) {$K$};
  \node[text=resgray!90!black,font=\footnotesize,inner sep=0pt]
   at (\x+.64,1.43) {$L$};
 }
 \draw[resblue!65,dash pattern=on 2pt off 1.5pt,line width=.55pt]
  (0,.15)--(0,1.45);
 \draw[resgray!90,line width=.8pt]
  (-.46,1.285)--(-.18,.80)--(-.46,.315)
  (.46,1.285)--(.18,.80)--(.46,.315);
 \draw[resblue,line width=.95pt] (-.18,.80)--(.18,.80);
 \node[text=resblue,font=\footnotesize,inner sep=0pt] at (.14,1.19) {$e$};
 \draw[resrust!65,dash pattern=on 2pt off 1.5pt,line width=.55pt]
  (1.75,.80)--(3.25,.80);
 \draw[resgray!90,line width=.8pt]
  (2.015,1.26)--(2.5,.98)--(2.985,1.26)
  (2.015,.34)--(2.5,.62)--(2.985,.34);
 \draw[resrust,line width=.95pt] (2.5,.62)--(2.5,.98);
 \node[text=resrust,font=\footnotesize,inner sep=0pt] at (2.83,.93) {$f$};
 \draw[<->,resgray,line width=.55pt] (.96,.80)--(1.54,.80);
 \node[text=resgray,font=\scriptsize,inner sep=0pt] at (1.25,1.06) {flip};
 \node[inner sep=0pt] (leftres) at (0,-.23) {$R^T_{m,\boldsymbol h}$};
 \node[inner sep=0pt] at (1.25,-.23) {$=$};
 \node[inner sep=0pt] at (2.5,-.23) {$R^{T'}_{r,\boldsymbol h},$};
\end{tikzpicture}%

%% file: figures/channel_flip.tex
\begin{tikzpicture}[x=1cm,y=1cm,>=Latex,
 every node/.style={font=\footnotesize},
 line cap=round,line join=round]
 \path[use as bounding box] (0,-7.32) rectangle (8.20,1.69);
 \node[anchor=west,font=\scriptsize\bfseries] at (.10,1.57) {(a)};

 \begin{scope}[shift={(1.83,0)}]
  \foreach \i/\ang in {1/90,2/18,3/-54,4/-126,5/162} {
   \coordinate (l\i) at (\ang:1.08);
   \node[font=\scriptsize,text=black!75] at (\ang:1.29) {\i};
  }
  \fill[resblue!7] (l1)--(l2)--(l3)--cycle;
  \draw[black!55,line width=.65pt]
   (l1)--(l2)--(l3)--(l4)--(l5)--cycle;
  \draw[resblue!70,line width=.85pt] (l1)--(l2)--(l3);
  \draw[resgray,line width=.95pt] (l1)--(l4);
  \draw[resblue,line width=1.15pt] (l1)--(l3);
  \node[text=resgray,font=\scriptsize] at (-.57,.15) {$14$};
  \node[text=resblue,font=\scriptsize] at (.57,.15) {$13$};
  \node[text=resblue,font=\scriptsize] at (.66,.91) {$p_1$};
  \node[text=resblue,font=\scriptsize] at (1.10,-.32) {$p_2$};
  \node at (0,1.57) {$T=(13,14)$};
  \node[text=resblue] at (0,-1.56)
   {$X_{13}=(p_1+p_2)^2$};
 \end{scope}

 \draw[<->,resgray!85,line width=.90pt] (3.52,.31)--(4.68,.31);
 \node[text=resgray,font=\scriptsize] at (4.10,.63) {flip};
 \node[align=center,text=resgray,font=\scriptsize]
  at (4.10,-.39) {common\\[-1pt]$X_{14}=-h$};

 \begin{scope}[shift={(6.37,0)}]
  \foreach \i/\ang in {1/90,2/18,3/-54,4/-126,5/162} {
   \coordinate (r\i) at (\ang:1.08);
   \node[font=\scriptsize,text=black!75] at (\ang:1.29) {\i};
  }
  \fill[resrust!7] (r2)--(r3)--(r4)--cycle;
  \draw[black!55,line width=.65pt]
   (r1)--(r2)--(r3)--(r4)--(r5)--cycle;
  \draw[resrust!70,line width=.85pt] (r2)--(r3)--(r4);
  \draw[resgray,line width=.95pt] (r1)--(r4);
  \draw[resrust,line width=1.15pt] (r2)--(r4);
  \node[text=resgray,font=\scriptsize] at (-.57,.15) {$14$};
  \node[text=resrust,font=\scriptsize] at (.25,-.55) {$24$};
  \node[text=resrust,font=\scriptsize] at (1.10,-.32) {$p_2$};
  \node[text=resrust,font=\scriptsize] at (0,-1.12) {$p_3$};
  \node at (0,1.57) {$T'=(24,14)$};
  \node[text=resrust] at (0,-1.56)
   {$X_{24}=(p_2+p_3)^2$};
 \end{scope}

 \node[font=\scriptsize,text=black!80] at (4.10,-2.03)
  {$\color{resblue}R^{13,14}_{m,h}
    \color{black!80}=\color{resrust}R^{24,14}_{r,h}
    \color{black!80}\qquad(c_{13}=-k,\ m+r=h+k)$};
 \draw[resgray!25,thin] (.30,-2.28)--(7.90,-2.28);
 \node[font=\scriptsize,text=black!80] at (4.10,-2.49)
  {Bootstrap at $(X_{24},X_{35},c_{24})=(-1,-1,-2)$:};
 \node[font=\scriptsize,text=resblue] at (1.00,-3.04)
  {$(13,14)_{1,1}$};
 \node[font=\scriptsize,text=black!80] at (3.83,-3.04)
  {$(24,14)_{1,1}$};
 \node[font=\scriptsize,text=resrust] at (6.87,-3.04)
  {$(24,25)_{1,0},\ R=2$};
 \draw[->,resgray,line width=.75pt] (1.87,-3.04)--(2.95,-3.04);
 \node[font=\scriptsize,text=resgray] at (2.41,-2.81)
  {$c_{13}=-1$};
 \draw[->,resgray,line width=.75pt] (4.71,-3.04)--(5.62,-3.04);
 \node[font=\scriptsize,text=resgray] at (5.17,-2.81)
  {$c_{14}=0$};

 \draw[resgray!25,thin] (.30,-3.43)--(7.90,-3.43);
 \node[anchor=west,font=\scriptsize\bfseries] at (.10,-3.69) {(b)};
 \node[font=\scriptsize,text=black!80] at (4.10,-3.69)
  {Six-point evaluation path: $X_{15}=-1$ (gray)};

 \begin{scope}[shift={(1.15,-4.96)}]
  \foreach \i/\ang in {1/90,2/30,3/-30,4/-90,5/-150,6/150} {
   \coordinate (s\i) at (\ang:.83);
   \node[font=\scriptsize,text=black!75] at (\ang:1.03) {\i};
  }
  \draw[black!55,line width=.65pt]
   (s1)--(s2)--(s3)--(s4)--(s5)--(s6)--cycle;
  \draw[resgray,line width=.95pt] (s1)--(s5);
  \draw[resblue,line width=1.05pt] (s1)--(s3) (s1)--(s4);
  \node[font=\scriptsize,text=black!80] at (0,-1.31)
   {$(13,14,15)_{1,1,1}$};
 \end{scope}

 \draw[->,resgray,line width=.75pt] (2.24,-4.96)--(3.01,-4.96);
 \node[font=\scriptsize,text=resgray] at (2.625,-4.71)
  {$13_1\to24_1$};
 \node[font=\scriptsize,text=resgray] at (2.625,-5.22)
  {$c_{13}=-1$};

 \begin{scope}[shift={(4.10,-4.96)}]
  \foreach \i/\ang in {1/90,2/30,3/-30,4/-90,5/-150,6/150} {
   \coordinate (t\i) at (\ang:.83);
   \node[font=\scriptsize,text=black!75] at (\ang:1.03) {\i};
  }
  \draw[black!55,line width=.65pt]
   (t1)--(t2)--(t3)--(t4)--(t5)--(t6)--cycle;
  \draw[resgray,line width=.95pt] (t1)--(t5);
  \draw[resrust,line width=1.05pt] (t2)--(t4);
  \draw[resblue,line width=1.05pt] (t1)--(t4);
  \node[font=\scriptsize,text=black!80] at (0,-1.31)
   {$(24,14,15)_{1,1,1}$};
 \end{scope}

 \draw[->,resgray,line width=.75pt] (5.19,-4.96)--(5.96,-4.96);
 \node[font=\scriptsize,text=resgray] at (5.575,-4.71)
  {$14_1\to25_0$};

 \begin{scope}[shift={(7.05,-4.96)}]
  \foreach \i/\ang in {1/90,2/30,3/-30,4/-90,5/-150,6/150} {
   \coordinate (u\i) at (\ang:.83);
   \node[font=\scriptsize,text=black!75] at (\ang:1.03) {\i};
  }
  \fill[resrust!9] (u2)--(u3)--(u4)--(u5)--cycle;
  \fill[resblue!7] (u1)--(u2)--(u5)--(u6)--cycle;
  \draw[black!55,line width=.65pt]
   (u1)--(u2)--(u3)--(u4)--(u5)--(u6)--cycle;
  \draw[resgray,line width=.95pt] (u1)--(u5);
  \draw[resrust,line width=1.05pt] (u2)--(u4);
  \draw[resrust,line width=1.35pt] (u2)--(u5);
  \node[font=\scriptsize,text=black!80] at (0,-1.31)
   {$(24,{\color{resrust}25},15)_{1,{\color{resrust}0},1}$};
 \end{scope}
 \node[font=\scriptsize,text=black!80] at (4.10,-6.72)
  {Evaluation points: $(X_{35},c_{24})=(-1,-2),\ (0,-1)$};
 \node[font=\scriptsize,text=black!80] at (4.10,-7.16)
  {Endpoint residue: $(1-X_{35})(1-X_{26})=4,\ 2$};
\end{tikzpicture}

%% file: figures/pinch_geometry.tex
\begin{tikzpicture}[x=1cm,y=1cm,>=Latex,
 every node/.style={font=\footnotesize},line cap=round,line join=round]
 \path[use as bounding box] (0,-2.03) rectangle (17.20,2.79);

 \node[anchor=west,font=\footnotesize\bfseries] at (0,2.57)
  {(a) One-dimensional pairing};
 \draw[resgray!50,thin,->] (.20,1.04)--(4.33,1.04);
 \draw[resblue!25,line width=4pt] (.91,1.04)--(3.43,1.04);
 \fill[resblue] (.91,1.04) circle (2.6pt);
 \fill[resrust] (3.43,1.04) circle (2.6pt);
 \node[text=resblue,font=\scriptsize] at (.91,1.55) {$\ell_e=w$};
 \node[text=resrust,font=\scriptsize] at (3.43,1.55)
  {$\ell_f=\Delta-w$};
 \node[text=resgray,font=\scriptsize] at (.91,.70) {$0$};
 \node[text=resgray,font=\scriptsize] at (3.43,.70) {$\Delta$};
 \node[anchor=west,text=resgray,font=\scriptsize] at (4.06,.70)
  {$\operatorname{Re}w$};
 \draw[resgray,line width=.90pt,->] (2.17,.53)--(2.17,2.17);
 \node[anchor=east,text=resgray,font=\scriptsize] at (1.98,.70)
  {$\sigma$};
 \node[anchor=west,text=resgray,font=\scriptsize] at (2.41,2.06)
  {$w=\sigma+it$};
 \node at (2.23,.11)
  {$\displaystyle F(w)=\frac{r_e}{w}+\frac{r_f}{\Delta-w}$};
 \node at (2.23,-.68)
  {$\displaystyle F(w)=\frac{r_e-r_f}{w}+O(w^{-2})$};
 \node[text=resgray,font=\scriptsize] at (2.23,-1.30)
  {absolute convergence};
 \node at (2.23,-1.70) {$\Longrightarrow\quad r_e=r_f$};

 \draw[resgray!25,thin] (4.96,-1.80)--(4.96,2.36);

 \node[anchor=west,font=\footnotesize\bfseries] at (5.36,2.57)
  {(b) Resolved five-point resonance};
 \begin{scope}[shift={(7.05,-.73)},x=1.45cm,y=.93cm]
  \fill[resblue!3] (0,-1)--(3,2)--(0,2)--cycle;
  \fill[resblue!11] (0,0)--(1,0)--(2,1)--(2,2)--(0,2)--cycle;
  \draw[resblue!45,densely dashed,line width=.65pt]
   (0,0)--(0,-1)--(1,0) (2,1)--(3,2)--(2,2);
  \draw[resblue,line width=1.05pt] (0,0)--(0,2)--(2,2);
  \draw[resblue,line width=1.05pt] (1,0)--(2,1);
  \draw[resrust,line width=1.15pt] (0,0)--(1,0) (2,1)--(2,2);
  \foreach \x/\y in {0/0,1/0,2/1,2/2,0/2}
   \fill[resgray] (\x,\y) circle (1.8pt);
  \node[rotate=90,text=resblue,font=\scriptsize] at (-.19,1.04)
   {$\ell_{13}=0$};
  \node[text=resblue,font=\scriptsize] at (1,2.26)
   {$\ell_{35}=0$};
  \node[rotate=33,text=resblue,font=\scriptsize] at (1.72,.33)
   {$\ell_{24}=0$};
  \node[text=resrust,font=\scriptsize] at (.48,.20)
   {$\ell_{14}=0$};
  \node[rotate=90,text=resrust,font=\scriptsize] at (1.80,1.50)
   {$\ell_{25}=0$};
  \node[anchor=east,font=\scriptsize] at (-.10,-.07) {$\{13,14\}$};
  \node[anchor=north,font=\scriptsize] at (1.10,-.25) {$\{14,24\}$};
  \node[anchor=west,font=\scriptsize] at (2.20,.95) {$\{24,25\}$};
  \node[anchor=south west,font=\scriptsize] at (2.07,2.06) {$\{25,35\}$};
  \node[anchor=south east,font=\scriptsize] at (-.07,2.06) {$\{13,35\}$};
 \end{scope}

 \node[anchor=west,text=black!80,font=\scriptsize] at (12.48,1.98)
  {$X_e=-m_e+\lambda\ell_e(W)$};
 \node[anchor=west,text=resgray,font=\scriptsize] at (12.48,1.45)
  {$\epsilon_{13}=\epsilon_{14}=\epsilon_{24}=1$};
 \draw[resblue,line width=1.05pt] (12.50,.92)--(12.97,.92);
 \node[anchor=west,font=\scriptsize] at (13.10,.92) {channels $13,24,35$};
 \draw[resrust,line width=1.15pt] (12.50,.49)--(12.97,.49);
 \node[anchor=west,font=\scriptsize] at (13.10,.49) {channels $14,25$};
 \node[anchor=west,font=\scriptsize] at (12.48,-.10)
  {$\ell_{13}+\ell_{24}+\ell_{35}=3$};
 \node[anchor=west,align=left,text=resgray,font=\scriptsize] at (12.48,-.86)
  {Five triangulations,\\one common residue $R_*$.};
 \node[anchor=west,text=black!80,font=\scriptsize] at (12.48,-1.63)
  {All $m_e=1$: $R_*=3$ (Table I).};
\end{tikzpicture}%

%% file: dual_resonance_pinch_supplement.tex
\section{Supplemental Material}
\setcounter{equation}{0}
\renewcommand{\theequation}{S\arabic{equation}}
\setcounter{figure}{0}
\renewcommand{\thefigure}{S\arabic{figure}}
\renewcommand{\theHfigure}{S\arabic{figure}}

Sections I--II derive the resonance relations, III--IV classify the
residues, V reconstructs the amplitude, and VI checks the string solution.

\paragraph{Notation and scope.}
We assume joint meromorphy in the independent planar invariants and the
compatible simple poles specified in the Letter. Residues are taken at
generic kinematics before imposing \(c_{ij}=-k_{ij}\); these conditions
do not introduce new poles. Channel labels \(ij\) abbreviate \((i,j)\),
with \(X_{i,i+1}=m_{i,i+1}=0\). Residue levels follow the displayed
channel order; fixed labels are suppressed. The superscript \({\rm str}\)
denotes the string solution. ADR is imposed at the two-pole collisions
of Section I and the complete integer resonances with bounded positive
ABHY region in Section II.

\subsection{I. Pairing under a flip}
\refstepcounter{smapp}\label{sm:I}
\label{app:fourpoint}

For a flip between \(T=S\cup\{e\}\) and \(T'=S\cup\{f\}\), write
\(e=(\alpha,\gamma)\), \(f=(\beta,\delta)\) and order the cyclic
vertex labels as \(\alpha<\beta<\gamma<\delta<\alpha+n\).
The two opposite pairs of particle groups give
\begin{equation}
 Q_+=[\alpha,\beta)\times[\gamma,\delta),\qquad
 Q_-=[\beta,\gamma)\times[\delta,\alpha+n).
 \label{app:cdef}
\end{equation}
Here \([\alpha,\beta)=\{\alpha,\ldots,\beta-1\}\), and indices are reduced cyclically.
Summing the ABHY relations gives
\begin{equation}
 \sum_{Q_+}c_{ij}=X_e+X_f-X_{\alpha\delta}-X_{\beta\gamma},\qquad
 \sum_{Q_-}c_{ij}=X_e+X_f-X_{\alpha\beta}-X_{\gamma\delta}.
 \label{app:rectangle}
\end{equation}
Take residues in the common channels \(S\), at levels \(\boldsymbol h\).
Choose either rectangle and set each \(c_{ij}=-k_{ij}\),
with integer \(k_{ij}\ge0\). Then
\(h_Q=m_{\alpha\delta}+m_{\beta\gamma}\) for \(Q_+\), or
\(m_{\alpha\beta}+m_{\gamma\delta}\) for
\(Q_-\), counting polygon sides as level zero. At
\(m+r=h_Q+\sum_Qk_{ij}\), the remaining poles meet at
\(X_e=-m+u\), \(X_f=-r-u\). Their \(1/u\) coefficient is the
difference of the two maximal residues. The two-pole ADR argument
in Eq.~\eqref{eq:local-period-tail} therefore gives
\begin{equation}
 \left.R^{S\cup\{e\}}_{(m,\boldsymbol h)}\right|_Q
 =\left.R^{S\cup\{f\}}_{(r,\boldsymbol h)}\right|_Q,
 \quad
 c_{ij}=-k_{ij},\quad k_{ij}\ge0,\quad
 \sum_Q k_{ij}=m+r-h_Q,\quad m,r\ge0.
 \label{app:physical-pair}
\end{equation}
The restrictions on each side are \(X_e=-m\), \(X_f=-r\), and the
displayed cell conditions. Only one rectangle is imposed at a time;
all remaining independent variables stay generic.

Ordinary duality alone allows satellites~\cite{GrossSatellites}, such as
\(\A_{\rm sat}=B(X_{13},X_{24})+\alpha_{\rm sat}B(X_{13}+1,X_{24}+1)\),
with \(\alpha_{\rm sat}\ne0\), where
\(B(x,y)=\Gamma(x)\Gamma(y)/\Gamma(x+y)\).
It has the Veneziano spectrum, degree bound, and leading fixed-\(X_{24}\)
Regge behavior, and admits both unsubtracted channel expansions.
At \(X_{13}=-1,X_{24}=0\), however, the two residues are
\(1+\alpha_{\rm sat}\) and \(1\): pairing fails.

At four points, the truncation zeros in Eq.~\eqref{eq:pair-zero},
\(\deg R_m\le m\), and \(R_m(-m)=R_0=1\) fix the residues and
the unsubtracted amplitude. These zeros do not follow from two-pole
ADR and are not imposed in our higher-point bootstrap.

\subsection{II. Simultaneous collisions}
\refstepcounter{smapp}\label{sm:II}
\label{app:contour-mechanism}
\label{app:adr-definition}

\paragraph{The complete pole sum.}
At a complete integer resonance, resolve all planar poles:
\begin{equation}
 X_e=-m_e+\lambda\ell_e(W),\qquad
 c_{ij}=-k_{ij}+\lambda\epsilon_{ij},\qquad
 m_e,k_{ij}\ge0,\quad \epsilon_{ij}>0 .
 \label{app:resonance-scaling}
\end{equation}
The fixed cells have \(1\le i<j\le n-1\), \(j-i\ge2\).
The levels and displacements obey the ABHY relations, including
Eq.~\eqref{eq:six-physical-node} at six points. In \(D=n-3\) fan
coordinates \(W_a=\ell_{1,a+2}\), the region \(\ell_e>0\) is the
bounded associahedron \(P\); choose \(\boldsymbol\sigma\) inside it.
Simple poles and vanishing incompatible multiple residues leave at
most \(D\) inverse channel factors in each local Laurent term.
Only triangulations therefore survive at order \(\lambda^{-D}\):
\begin{equation}
 F(W)=\lim_{\lambda\to0}\lambda^D\A_n
       (-\boldsymbol m+\lambda\boldsymbol\ell,
        -\boldsymbol k+\lambda\boldsymbol\epsilon)
     =\sum_T\frac{r_T}{\prod_{e\in T}\ell_e(W)} .
 \label{app:leading-block}
\end{equation}
Here \(r_T\) is the generic maximal residue evaluated at the integer
node. It is regular there because every additional diagonal crosses
an edge of \(T\). As required by ADR, Eq.~\eqref{eq:adr}, the limit
precedes integration.

\paragraph{All colliding channels at once.}
\label{app:adr-allrank}

ADR holds precisely when all \(r_T\) agree.
Write \(\ell_e(\boldsymbol\sigma+it)
=\ell_e(\boldsymbol\sigma)+i\boldsymbol\nu_e\cdot t\).
The incident normals at each vertex \(W_T\) have unit absolute
determinant: this holds for the coordinate fan, and a flip replaces
\(\boldsymbol\nu_e\) by \(-\boldsymbol\nu_e+
\boldsymbol\nu_{\alpha\delta}+\boldsymbol\nu_{\beta\gamma}\).
The added rows are common or zero, so only the determinant's sign changes;
flip connectivity proves the claim for every vertex.
Apply \(1/z=\int_0^\infty e^{-\beta z}\dd\beta\) to each pole
factor and set \(v=\sum_e\beta_e\boldsymbol\nu_e\), with unit
absolute Jacobian. Positive \(\beta_e\) fill the inward normal cone
\(C_T\) at \(W_T\), giving
\begin{equation}
 F(\boldsymbol\sigma+it)=\int_{\mathbb R^D}\dd^Dv\,\rho(v)e^{-it\cdot v},
 \qquad
 \rho(v)=r_Te^{-(\boldsymbol\sigma-W_T)\cdot v}\quad(v\text{ inside }C_T).
 \label{app:resonance-fourier}
\end{equation}
These cones partition \(\mathbb R^D\) up to their boundaries.
Since \(\boldsymbol\sigma\) is interior, \(\rho\) decays exponentially.
ADR makes its Fourier transform absolutely integrable, so inverse
transformation gives a continuous function agreeing with the displayed
exponential inside every cone. Approaching the origin from any cone
therefore gives the same value: \(r_T=R_*\).

Conversely, for \(r_T=R_*\) the pole sum is \(R_*\) times the
canonical function of \(P\)~\cite{ABHY}. Triangulate \(P\) with
facet hyperplanes avoiding \(\boldsymbol\sigma\), adding generic
vertices if needed. Each simplex contributes a constant numerator over
barycentric denominators \(L_0,\ldots,L_D\), with
\(\sum_iL_i=1\)~\cite{PositiveGeometry}. Their real parts are
nonzero. Using the imaginary parts of \(L_1,\ldots,L_D\) as
coordinates bounds each absolute integral by a constant times
\begin{equation}
 \int_{\mathbb R^D}
 \frac{\dd^Dt}{(1+|t_1|)\cdots(1+|t_D|)
                    (1+|t_1+\cdots+t_D|)}<\infty .
 \label{app:simplex-bound}
\end{equation}
This integral is the convolution of \(D+1\) copies of
\((1+|t|)^{-1}\) evaluated at zero. It is finite by Young's inequality,
since this function lies in \(L^{(D+1)/D}\).
Fourier inversion at the origin then gives
\begin{equation}
 \int_{\boldsymbol\sigma+i\mathbb R^D}\frac{\dd^DW}{(2\pi i)^D}F(W)=R_* .
 \label{app:resonance-period}
\end{equation}
\label{app:physical-circuit}
The same equalities follow from paired flips. For
\(1\le\alpha<\beta<\gamma<\delta\le n\), the rectangle \(Q_+\)
lies in the fixed cell set. Equation~\eqref{app:rectangle} gives
\(m_e+m_f=h_Q+\sum_{Q_+}k_{ij}\), so every flip is permitted.
Flip connectivity equates all maximal residues. The simultaneous
collision is a different limit, but requires no independent bootstrap input.

\paragraph{The complete five-point collision.}
\label{app:five-overlap}

An independent five-point check uses the beta-series expansion
of the known hypergeometric representation~\cite{OpenStringModern}:
\begin{equation}
 \A_5^{\rm str}
 =\sum_{j\ge0}\frac{(c_{24})_j}{j!}
 B(X_{13}+j,X_{24})B(X_{14}+j,X_{35})\,,
 \label{app:five-hypergeometric}
\end{equation}
The series converges initially for \(\Re X_{25}>0\), away from
its parameter poles, and is then continued meromorphically.
At an allowed integer resonance, let \(R_*\) be
Eq.~\eqref{eq:five-all} evaluated at \(a=m_{13}\), \(b=m_{14}\),
\(X_{24}=-m_{24}\), \(X_{35}=-m_{35}\), and \(c_{24}=-k_{24}\).
The ABHY relation gives
\(k_{24}=m_{24}+m_{35}-m_{25}\) and
\(\epsilon_{24}=\ell_{24}+\ell_{35}-\ell_{25}\).
For \(j\le m_{13}\), the first beta factor has leading term
\(\lambda^{-1}\binom{m_{13}+m_{24}-j}{m_{24}}
(\ell_{13}^{-1}+\ell_{24}^{-1})\); the second has the analogous
form with \(13,24\) replaced by \(14,35\).
Multiplying by \((-k_{24})_j/j!=(-1)^j\binom{k_{24}}j\) and
summing gives precisely the coefficient in Eq.~\eqref{eq:five-all}.
Thus the finite subsum \(0\le j\le k_{24}\) gives
\begin{equation}
 \lambda^2\A_{5,\mathrm{finite}}^{\rm str}
 \longrightarrow R_*
 \left(\frac1{\ell_{13}}+\frac1{\ell_{24}}\right)
 \left(\frac1{\ell_{14}}+\frac1{\ell_{35}}\right).
 \label{app:five-finite-leading}
\end{equation}
Allowed integer levels make terms with
\(j>m_{13}\) or \(j>m_{14}\) subleading.

The infinite tail \(j>k_{24}\) cannot be discarded term by term.
Write its exact summand as \(\mathcal N(\lambda)\mathcal K_j(\lambda)\), where
\begin{equation}
 \begin{aligned}
 \mathcal N(\lambda)&=\frac{\Gamma(X_{24})\Gamma(X_{35})}{\Gamma(c_{24})}
 =\frac{(-1)^{m_{25}}k_{24}!\epsilon_{24}}
 {m_{24}!m_{35}!\lambda\ell_{24}\ell_{35}}+O(1),\\
 \mathcal K_j(\lambda)&=\frac{\Gamma(j+c_{24})}{\Gamma(j+1)}
 \frac{\Gamma(j+X_{13})}{\Gamma(j+X_{13}+X_{24})}
 \frac{\Gamma(j+X_{14})}{\Gamma(j+X_{14}+X_{35})}.
 \end{aligned}
 \label{app:tail-summand}
\end{equation}
All arguments follow Eq.~\eqref{app:resonance-scaling}.
The large-\(j\) expansion has the form
\begin{equation}
 \mathcal K_j(\lambda)=\sum_{s=0}^{m_{25}}A_s(\lambda)
 j^{m_{25}-1-s-\lambda\ell_{25}}
 +O\!\left(j^{-2-\Re(\lambda\ell_{25})}\right).
 \label{app:tail-subtractions}
\end{equation}
Subtracting these \(m_{25}+1\) terms leaves a locally uniformly
convergent sum for \(|\Re(\lambda\ell_{25})|<1/2\).
The subtracted powers sum to Riemann zeta functions
\(\zeta(1-m_{25}+s+\lambda\ell_{25})\), with the first
\(k_{24}\) terms removed. Only \(s=m_{25}\) is singular at
\(\lambda=0\), and it contributes
\(A_{m_{25}}(0)/(\lambda\ell_{25})\).

The large-\(j\) expansion at \(\lambda=0\) is that of the rational function
\begin{equation}
 \mathcal U(z)=\frac{(z-m_{13}-m_{24})_{m_{24}}
 (z-m_{14}-m_{35})_{m_{35}}}{(z-k_{24})_{k_{24}+1}},\qquad
 A_{m_{25}}(0)=[z^{-1}]_{\infty}\mathcal U(z)
 =\sum_{j=0}^{k_{24}}\Res_{z=j}\mathcal U(z).
 \label{app:tail-rational}
\end{equation}
The bracket extracts the \(1/z\) coefficient at large \(z\).
The allowed-level inequalities \(m_{14}\le m_{13}+m_{24}\)
and \(k_{24}\le m_{14}+m_{35}\) make these residues vanish when
\(j>\min(m_{13},m_{14})\). For the remaining terms, expressing the
rising factorials as binomial coefficients and comparing with
Eq.~\eqref{eq:five-all} gives
\[
 A_{m_{25}}(0)=\frac{(-1)^{m_{25}}m_{24}!m_{35}!}{k_{24}!}R_*.
\]
Multiplying the singular zeta term by \(\mathcal N(\lambda)\)
therefore yields
\begin{equation}
 \lambda^2\A_{5,\mathrm{tail}}^{\rm str}
 \longrightarrow\frac{R_*\epsilon_{24}}
 {\ell_{24}\ell_{35}\ell_{25}}.
 \label{app:five-tail-leading}
\end{equation}
Adding the finite and continued tail contributions gives
\begin{equation}
 \lim_{\lambda\to0}\lambda^2\A_5^{\rm str}
 =R_*\left(
 \frac1{\ell_{13}\ell_{14}}+\frac1{\ell_{14}\ell_{24}}
 +\frac1{\ell_{24}\ell_{25}}+\frac1{\ell_{25}\ell_{35}}
 +\frac1{\ell_{35}\ell_{13}}\right).
 \label{app:five-full-overlap}
\end{equation}
The tail cancels the apparent incompatible double pole \((24,35)\) and supplies
the two terms containing \(25\), independently verifying the complete
collision at all allowed levels.

\subsection{III. Five points and the mass-level induction}
\refstepcounter{smapp}\label{sm:III}
\label{app:inputs}

Pairing and the joint degree bound, Eq.~\eqref{eq:bootstrap}, suffice.
We first fix the five-point zero-level residues, then prove an induction
for positive levels at any \(n\). Section IV supplies the higher-point
zero-level data; factorization is not assumed.

\paragraph{Five-point residues with one zero level.}
\label{app:five-scalar-free}

Write the five independent maximal-residue families as
\(F_i(a,b;p,q,\chi)\), with \(i=0,\ldots,4\) understood modulo five.
Their ordered channels and local kinematic coordinates are
\begin{equation}
 \begin{array}{c|c|c|c|l}
 i & F_i(a,b;p,q,\chi) & p & q & \chi \\ \hline
 0 & R^{13,14}_{a,b} & X_{24} & X_{35}
   & c_{24}=X_{24}+X_{35}-X_{25} \\[2pt]
 1 & R^{14,24}_{a,b} & X_{25} & X_{13}
   & c_{25}=X_{25}+X_{13}-X_{35} \\[2pt]
 2 & R^{24,25}_{a,b} & X_{35} & X_{14}
   & c_{35}=X_{35}+X_{14}-X_{13} \\[2pt]
 3 & R^{25,35}_{a,b} & X_{13} & X_{24}
   & c_{13}=X_{13}+X_{24}-X_{14} \\[2pt]
 4 & R^{35,13}_{a,b} & X_{14} & X_{25}
   & c_{14}=X_{14}+X_{25}-X_{24} .
 \end{array}
 \label{app:five-families}
\end{equation}
Adjacent rows are related by a flip. Relabeling all external vertices
by \(r\mapsto r+3\) modulo five advances one row; it does not identify
the five unknown functions or assume cyclic symmetry.

In each chart, the joint degree bound reads
\(\deg_{p,\chi}F_i\le a\), \(\deg_{q,\chi}F_i\le b\).
Thus \(F_i(a,0;x)\) has only the argument \(x=p\), and
\(F_i(0,b;x)\) only \(x=q\). For example, at
\(X_{14}=0,X_{13}=-a,X_{24}=-d\), the flip
\(13\leftrightarrow24\) has \(c_{13}=-(a+d)\le0\).
Equation~\eqref{app:physical-pair} and its relabelings give
\begin{equation}
 G_i(a,d):=F_i(a,0;-d)=F_{i+1}(0,d;-a),\qquad a,d\ge0.
 \label{app:five-zero-pair}
\end{equation}
The first expression has degree at most \(a\) in \(d\), the second
degree at most \(d\) in \(a\). Expand in binomials
\(\binom a\ell\binom dj\). Their coefficients are obtained by
taking \(\ell\) forward differences in \(a\) and \(j\) in \(d\)
at the origin. If \(\ell>j\), the coefficient vanishes:
the finite difference in \(d\) samples only \(d=0,\ldots,j\), where
the degree in \(a\) is below \(\ell\). Interchanging \(a,d\)
also excludes \(j>\ell\). Only diagonal terms remain, so
\(G_i(a,d)=G_i(d,a)\). Polynomial continuation identifies
\(F_i(a,0;x)=F_{i+1}(0,a;x)=\varphi_{i,a}(x)\), with
\begin{equation}
 \varphi_{i,a}(x)=\sum_{j=0}^a A_{i,j}\binom aj\binom{-x}j,
 \qquad A_{i,0}=g,\qquad t_i=A_{i,1}.
 \label{app:five-axis}
\end{equation}
Here \(g\) is the common all-zero-level residue and
\(\varphi_{i,1}(x)=g-t_ix\). This is the four-point binomial
parameterization~\cite{StringyDynamics}, with coefficients still
independent between the five boundaries.

For \(a\ge1\), \(F_i(a,1)\) has total degree at most one in \(q,\chi\).
Take levels \((a,1,d,d-j,0)\), with \(j=0,1\), \(d\ge j\),
ordered as \((13,14,24,25,35)\) for \(i=0\) and relabeled for other \(i\).
The path \(i\to i+1\to i+2\to i+3\) uses single-cell rectangles
with successive values \(k=a+d-1,1-j,j\), all nonnegative, and ends at a zero level:
\begin{equation}
 F_i(a,1;-d,0,-j)=\varphi_{i+3,d-j}(-a)
                 =\varphi_{i+3,a}(-d+j).
 \label{app:five-path-values}
\end{equation}
The last equality uses the boundary symmetry just derived.
The infinitely many \(d\) values and linearity in \(\chi\) fix
\(F_i\) at \(q=0\). At \(p=0\), use \(13_a\to24_0\) with
spectator \(14_1\): \(c_{13}=1-a\le0\) and \(X_{25}=q-\chi\).
The two restrictions are
\begin{align}
 F_i(a,1;p,0,\chi)&=\varphi_{i+3,a}(p)
   +\chi\,[\varphi_{i+3,a}(p)-\varphi_{i+3,a}(p+1)],\nonumber\\
 F_i(a,1;0,q,\chi)&=\varphi_{i+1,1}(q-\chi)
                  =g-t_{i+1}q+t_{i+1}\chi.
 \label{app:five-axis-recurrence}
\end{align}
Comparing \(\chi\) coefficients at \(p=q=0\) gives
\(\varphi_{i+3,a}(0)-\varphi_{i+3,a}(1)=t_{i+1}\).
At \(a=1\), the comparison gives \(t_{i+3}=t_{i+1}\).
Shifts by two connect all five charts, so all \(t_i\) equal a common
\(t\). Hence \(\varphi_{i,a}(1)=g-t\) for \(a\ge1\). Since
\(\binom{-1}{j}=(-1)^j\), binomial inversion of
Eq.~\eqref{app:five-axis} gives \(A_{i,j}=t\) for every \(j\ge1\).
Therefore \(\varphi_{i,a}(x)=t(1-x)_a/a!+g-t\), with
\(t=t_5\) and \(q_5=g-t\).

\label{app:alln}

\paragraph{Induction on the total mass level.}
At fixed \(n\ge5\), let \(H^T_{\boldsymbol m}\) obey pairing and
the joint degree bound. We will apply the argument to
\(H^T_{\boldsymbol m}=R^T_{\boldsymbol m}
-t_nR^{T,\rm str}_{\boldsymbol m}-q_n\).
The constants \(t_n,q_n\) will be fixed by the zero-level data,
independently of levels and triangulation.
Throughout the proof, \(H=0\) means a polynomial identity at generic
remaining kinematics, not just a zero on a pairing locus.

We induct on the total level \(M=\sum_{e\in T}m_e\).
Assume all residues of total below \(M\) vanish, including those with
zero levels. We show that every all-positive residue at total \(M\)
vanishes, without using zero-level data at this total. Pairing first
forces polynomial factors, then vanishing in a five-point region;
allowed flips extend the result to every triangulation.

\paragraph{How a flip uses the degree bound.}
Consider \(T=S\cup\{e\}\) and \(T'=S\cup\{f\}\), with common
levels \(\boldsymbol h\). Write
\(H=H^T_{(m,\boldsymbol h)}\) and
\(H'_r=H^{T'}_{(r,\boldsymbol h)}\).
Replacing \(m\) by \(r\) changes the total to \(M-m+r\), so
induction gives \(H'_r=0\) for \(r<m\), but not for \(r=m\).
The two arguments below also apply when the required partners vanish
by another argument, as in Section IV.
Write \(e=(\alpha,\gamma)\), \(f=(\beta,\delta)\), with cyclically
ordered vertices \(\alpha<\beta<\gamma<\delta<\alpha+n\), and choose
one of the rectangles \(Q_+\) or \(Q_-\) in Eq.~\eqref{app:cdef}.
For \(Q_+=[\alpha,\beta)\times[\gamma,\delta)\), hold row
\(i=\alpha\) and column \(j=\gamma\) of \(X_{ij}\) fixed.
The ABHY relations determine \(X_{ij}\) with
\(\alpha<i\le\beta\), \(\gamma<j\le\delta\):
\begin{equation}
 X_{ij}=X_{\alpha j}+X_{i\gamma}-X_{\alpha\gamma}
 +\sum_{u=\alpha}^{i-1}\sum_{v=\gamma}^{j-1}c_{uv}.
 \label{app:rectangle-coordinates}
\end{equation}
These channels cross \(e\). Replacing them by the cell variables
is an invertible affine change preserving their total degree.
For \(Q_-\), start at row \(i=\gamma\), column \(j=\alpha+n\)
and proceed northwest. Other independent variables stay generic.

On the slice \(X_f=-r\), telescoping the cell equations leaves the
single relation
\begin{equation}
 \sum_{(i,j)\in Q}k_{ij}=N_Q=m+r-h_Q,
 \qquad k_{ij}\ge0.
 \label{app:flip-simplex}
\end{equation}
Pairing gives a value at every nonnegative integer
composition of \(N_Q\). If \(Q\) has \(s\) cells, eliminating one
cell leaves the integer simplex in \(s-1\) variables with coordinate
sum at most \(N_Q\). A polynomial of total degree at most \(N_Q\)
is determined by its values on this simplex: the products of binomial
polynomials of total order at most \(N_Q\) give a triangular
interpolation basis. For a single cell, the sum fixes its value completely.

Suppose \(H'_r=0\) for \(r<m\) and \(h_Q<m\).
Descend from \(r=m-1\) to zero. After removing the factors
\(\prod_{j=r+1}^{m-1}(X_f+j)\), the quotient has crossing degree
at most \(r+1\). The removed factors are nonzero at \(X_f=-r\),
where pairing makes the quotient vanish at every integer node.
Since \(N_Q\ge r+1\), interpolation extends this zero to the whole
hyperplane, giving a further factor \(X_f+r\). Thus
\begin{equation}
 h_Q<m\quad\Longrightarrow\quad H=(X_f)_m\widetilde H.
 \label{app:down-factor}
\end{equation}
The factor saturates the joint degree bound, so \(\widetilde H\)
is independent of every variable crossing \(e\). Since \(f\)
crosses no other edge of \(T\), the other degree bounds are unchanged.

If also \(H'_m=0\), start the descent at \(r=m\).
The remaining degree is now at most \(r\), so \(h_Q\le m\)
suffices. The resulting \(m+1\) factors exceed the degree bound:
\begin{equation}
 H_{\rm neighbor}=0,\quad h_Q\le m\quad\Longrightarrow\quad H=0.
 \label{app:transfer}
\end{equation}
Here \(H_{\rm neighbor}=H'_m\). This transfer rule requires all
lower partners to vanish as well; the inequality alone is not enough.

\paragraph{Two comparisons in a five-point region.}
Take a pentagon with consecutive vertices \(1,\ldots,5\), four
original sides of level zero, and a single attachment \(15\).
Let \(T\) contain \(13_a,14_b\), with \(a,b>0\), and obtain
\(T'\) by the same-level flip \(13_a\to24_a\).
Keep the attachment level \(m_{15}\) and all exterior data arbitrary;
at five points, \(15\) is a polygon side.

For the flip \(13\leftrightarrow24\), choose
\(Q=\{2\}\times\{4,\ldots,n\}\). Its opposite-side sum is
\(h_Q=m_{12}+m_{34}=0\), independently of \(m_{14}\) and
\(m_{15}\). The flip \(14\leftrightarrow25\) in \(T'\) similarly
uses \(Q=\{2,3\}\times\{5,\ldots,n\}\), with
\(h_Q=m_{12}+m_{45}=0\). Lowering either exchanged level gives a
partner of total less than \(M\), which vanishes by induction.
Equation~\eqref{app:down-factor} applies to \(13/24\) in both
directions and to \(14/25\) in \(T'\), giving
\begin{equation}
 H^T=(X_{24})_aU,\qquad
 H^{T'}=(X_{13})_a(X_{25})_bV.
 \label{app:seed-factors}
\end{equation}
Saturation of the \(13\) bound makes \(U\) independent of
\(X_{24},\ldots,X_{2n}\). In \(T'\), saturation of the \(24\) and
\(14\) bounds makes \(V\) independent of
\(X_{13},X_{25},\ldots,X_{2n},X_{35},\ldots,X_{3n}\).
Other independent variables and common levels remain unrestricted.

On the pairing locus at equal levels, \(X_{13}=X_{24}=-a\), the first
strip obeys \(\sum_{j=4}^n k_{2j}=2a\), and its individual equations give
\begin{equation}
 X_{2j}=X_{3j}-a+\sum_{q=4}^{j-1}k_{2q},\qquad 5\le j\le n.
 \label{app:seed-coordinates}
\end{equation}
Choose \((k_{24},\ldots,k_{2n})=(0,\ldots,0,2a)\) or
\((1,0,\ldots,0,2a-1)\), both allowed since \(a>0\).
Hold row-three and exterior variables fixed. Only row-two entries
change, so the quotients \(U,V\) are the same at both choices,
where \(X_{25}=X_{35}-a\) and \(X_{35}-a+1\), respectively.
Canceling \((-a)_a=(-1)^aa!\) in the paired residues gives
\begin{equation}
 U=(X_{35}-a)_bV=(X_{35}-a+1)_bV,
 \qquad b(X_{35}-a+1)_{b-1}V=0.
 \label{app:seed-zero}
\end{equation}
Since \(b>0\), this forces \(V=U=0\) at arbitrary attachment
level and generic exterior data.

\begin{figure}[htbp]
\centering
\input{figures/weighted_arms}
\caption{Allowed flips producing a terminal pentagon.
Solid lines show the triangulation; dashed links form the dual tree,
with blue dots for ears. (a--c) The short-arm configurations attach
to the exterior at level \(c\); other boundary sides have level zero.
Letters label mass levels. (d) For \(b\ge a\), the flip
\(36_b\to15_b\) obeys \(\min(a,c+d)\le b\) and produces
the shaded pentagon \(12345\), attached at level \(b\).}
\label{fig:weighted-arms}
\end{figure}
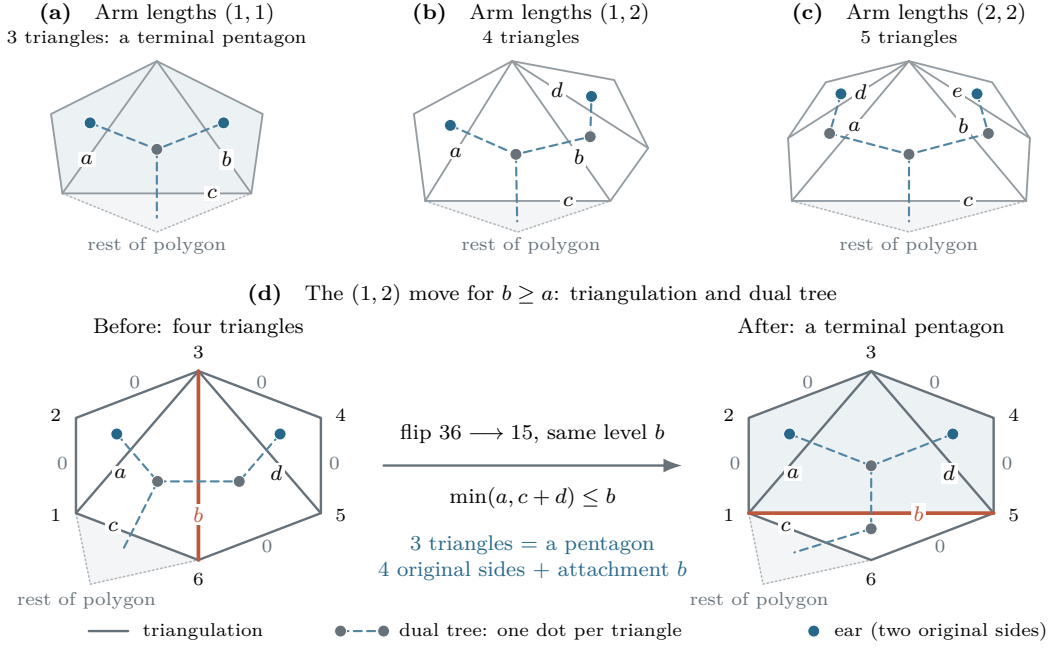

\paragraph{Reaching the same region from any triangulation.}
Give each diagonal its mass level as a weight, and polygon sides
weight zero. A same-level flip preserves \(M\).
For quadrilateral side weights \(a,b,c,d\) in cyclic order, either
rectangle can be used in Eq.~\eqref{app:transfer}. Vanishing can
therefore pass between same-level neighbors when
\begin{equation}
 \min(a+c,b+d)\le m.
 \label{app:weighted-flip}
\end{equation}
We construct such a path using the dual tree in
Fig.~\ref{fig:weighted-arms}. A terminal subtree of three triangles
is a pentagon with four original sides and one attachment. Root the
tree at an ear, a triangle with two original sides. A path, or an
outward arm of at least three triangles, already supplies this pentagon.
Otherwise the farthest branch has arm lengths \((1,1),(1,2)\),
or \((2,2)\), excluding the branch itself [panels (a--c)].
The \((1,1)\) case is immediate. A two-triangle arm can be reoriented
by an always-allowed flip: three of its quadrilateral sides are
original sides, so one rectangle has \(h_Q=0\).

For \((1,2)\), use the stem weights \(a,b\) in panel (b).
If \(a\ge b\), flipping \(a\) obeys \(\min(b,c)\le a\)
and turns the four triangles into a path. If \(b\ge a\), orient
the long arm as in panel (d), where \(a=m_{13},b=m_{36},
c=m_{16},d=m_{35}\). The flip \(36_b\to15_b\) obeys
\(\min(a,c+d)\le b\) and leaves the pentagon \(12345\)
attached to the exterior along \(15_b\).

For \((2,2)\), take \(a\ge b\) and orient the \(a\)-arm as in
panel (c). Flipping \(a\) obeys \(\min(b,c+d)\le a\).
It leaves a connector with sides \((c,0,a)\) and a \((1,2)\)
patch attached at \(a\), with short and long stems \(d,b\).
The preceding case now applies.

Label the pentagonal subpolygon's attachment diagonal \(15\), at arbitrary level
\(m_{15}\). Its triangulations are
\(\{13,14\}\), \(\{14,24\}\), \(\{24,25\}\),
\(\{25,35\}\), and \(\{35,13\}\).
The first and fourth are fans based at the two endpoints of the
attachment diagonal. The first is the configuration of
Eq.~\eqref{app:seed-factors}; the fourth follows by repeating the
argument with reversed local vertex labels. This uses pairing and the
degree bound, not reflection symmetry of the unknown residues.
The second triangulation reaches
the first by \(24\to13\), and the third the fourth by \(24\to35\),
both with \(h_Q=0\). In the last, let \(13,35\) have weights
\(x,y\). If \(x\ge y\), flip \(13\to25\), using
\(\min(y,m_{15})\le x\), to reach the fourth triangulation.
If \(y\ge x\), flip \(35\to14\), using
\(\min(x,m_{15})\le y\), to reach the first.

Follow this path backwards from a vanishing fan. Each same-level
neighbor is then known to vanish, and all lower partners vanish by
induction, so Eq.~\eqref{app:transfer} sets every preceding residue
to zero. The path retains positive levels throughout; no zero-level
data at total \(M\) have been used.

At five points, the earlier boundary calculation makes
\(H_i=F_i-tF_i^{\rm str}-(g-t)\) vanish whenever either level is
zero. Induction on \(M=a+b\) therefore gives
\(F_i=tF_i^{\rm str}+g-t\) for every chart and level.

\subsection{IV. Completing the higher-point proof}
\refstepcounter{smapp}\label{sm:IV}

\label{app:component-lift}

On \(X_{1p}=0\), the degree bound removes \(X_{ij}\) with
\(1<i<p<j\). The two polygons are \(P_L=(1,\ldots,p)\),
\(P_R=(1,p,p+1,\ldots,n)\). Fix the triangulation and levels
\(\boldsymbol h\) in \(P_R\), with remaining invariants \(\eta\)
generic. Lift a pairing inside \(P_L\) by specializing only
the removed variables:
\begin{equation}
 X_{ij}=X_{ip}\quad(1<i<p<j\le n),\qquad
 c_{ij}=0\ (p\le j<n),\qquad c_{in}=\bar c_{ip}.
 \label{app:scalar-lift}
\end{equation}
Here \(2\le i\le p-2\) in the last two equations, and the bar
denotes a cell computed within \(P_L\), where \(p\) is followed by \(1\):
\(\bar c_{ip}=X_{ip}+X_{i+1,1}-X_{i1}-X_{i+1,p}\).
For a flip
\(\alpha<\beta<\gamma<\delta\le p\), \(Q_+\) stays inside \(P_L\). In \(Q_-\),
all rows are strictly interior. Each cell at the closing side becomes
\(\bar c=-k\mapsto(0,\ldots,0,-k)\). The cell integers remain
nonnegative and the side-level sum is unchanged. Both rectangles give
lower-point pairing without restricting component kinematics.

The fixed levels in \(P_R\) bound the degree in every \(\eta\)
variable, independently of the levels in \(P_L\). Expand in this
common finite monomial basis. Each coefficient obeys \(p\)-point
pairing and degree bounds, so the classification at \(p\ge5\) gives
\begin{equation}
 R_n\big|_{m_{1p}=0}=A_{\boldsymbol h}(\eta)R_p^{\rm str}
                         +B_{\boldsymbol h}(\eta).
 \label{app:component-classification}
\end{equation}
The coefficients multiplying the residue family \(R_p^{\rm str}\)
are independent of its levels and triangulation, but may depend on \(P_R\).

When \(P_R\) is an ear, it has no levels or invariants, so the residue
is \(t_eR_{n-1}^{\rm str}+q_e\) with numerical coefficients.
Compatible ears share a double-zero
restriction proportional to \(R_{n-2}^{\rm str}\) plus a constant.
These families are independent at every multiplicity \(r\ge4\):
zero-level restrictions of the string solution reduce to four points,
whose residues at levels zero and one are \(1\) and \(1-x\),
whereas the constant family stays \(1\). Compatible ears therefore
have equal coefficients. Each ear is incompatible only with its
two neighbors, so compatible pairs connect all ears for \(n\ge6\).
All ears therefore share \(t_n,q_n\), without assuming cyclic symmetry.
Subtracting \(t_nR_n^{\rm str}+q_n\) makes residues on zero-level ears
vanish by the reference solution's level-zero limit.

\paragraph{The six-point split into two quadrilaterals.}
\label{app:six-scalar-free}

At six points, \(H=R_6-t_6R_6^{\rm str}-q_6\) vanishes on every
ear. Any other zero-level channel splits the hexagon into two quadrilaterals.
Choose a nonzero residue on these cuts minimizing the total level
\(M_\star=a+b\) over all component triangulations and levels.
Both levels are positive, since a second zero gives an ear restriction.
Relabel the chosen zero-level cut as \(25\), without identifying
different residue functions by symmetry.
Internal quadrilateral flips have \(h_Q=0\). If a same-level
neighbor vanished, minimality and \(a+1\) or \(b+1\) interpolation
nodes would force the original residue to vanish. Same-level flips
within the two quadrilateral components therefore give a nonzero
\(\{24,25,15\}_{a,0,b}\) residue at the same minimum, leaving the
zero-level cut \(25\) unchanged.

All zero-level residues below \(M_\star\) vanish. Section III then
sets every all-positive residue through \(M_\star\) to zero,
without needing zero-level data at the current total.
At total level \(M_\star+1\), consider
\(T=\{13,14,15\}_{a,1,b}\), \(T'=\{24,14,15\}_{a,1,b}\).
Use Fig.~\ref{fig:channels}(b) at levels \((a,1,b)\).
Lowering an outer level gives either an
all-positive residue of total at most \(M_\star\) or a zero-level ear
restriction. Both vanish; the unknown middle-zero residue at \(M_\star\)
is never used. Since both outer flips have \(h_Q=0\),
Eq.~\eqref{app:down-factor} gives
\begin{equation}
 H^T=(X_{24})_a(X_{46})_b(\alpha_0+\alpha_1 X_{35}),\qquad
 H^{T'}=(X_{13})_a(X_{46})_b(\beta_0+\beta_1 X_{25}).
 \label{app:six-bridge-shapes}
\end{equation}
The remaining middle crossing variable has degree at most one;
the coefficients are constant at fixed \(a,b\).

The \(13/24\) strip admits
\((k_{24},k_{25},k_{26})=(0,0,2a),(1,0,2a-1)\), with
\(X_{35},X_{36},X_{46}\) generic. Using
\(X_{25}=X_{35}-a+k_{24}\) and canceling common factors gives
\(\alpha_0+\alpha_1X_{35}=\beta_0+\beta_1(X_{35}-a)
=\beta_0+\beta_1(X_{35}-a+1)\).
Hence \(\alpha_1=\beta_1=0\), \(\alpha_0=\beta_0\).

Now flip \(14_1\to25_0\) in \(T'\) on
\(Q=\{2,3\}\times\{5,6\}\), where \(h_Q=0\) and \(\sum k=1\).
Keep \(u=X_{35}\), \(v=X_{26}\) generic. The choices \(k_{36}=1\)
and \(k_{26}=1\), with other cells zero, give respectively
\((X_{13},X_{46})=(u,v)\) and \((u-1,v)\).
The target's zero-level bound removes the changed crossing variables,
so both choices evaluate the same polynomial in \(u,v\):
\begin{equation}
 H^{\{24,25,15\}}_{a,0,b}(u,v)
 =\beta_0(u)_a(v)_b=\beta_0(u-1)_a(v)_b.
 \label{app:six-bridge-zero}
\end{equation}
Subtracting gives \(\beta_0a(u)_{a-1}(v)_b=0\), so \(\beta_0=0\),
a contradiction. All zero-level residues vanish; positive-level
induction then gives
\(R_6=t_6R_6^{\rm str}+q_6\).

\paragraph{Closing the multiplicity induction.}
\label{app:alln-scalar-free}

For \(n\ge7\), assume the classification for \(5\le p<n\).
Ear comparison makes \(H=R_n-t_nR_n^{\rm str}-q_n\) vanish
on every original ear. Any other zero-level channel splits the polygon
into two polygons of sizes \(p,q\), with \(p+q=n+2\); label them so
that \(5\le p<n\).
Equation~\eqref{app:component-classification} gives
\(H|_{m_e=0}=A(\eta)R_p^{\rm str}+B(\eta)\).
Set an ear channel common to this component and the original \(n\)-gon
to level zero, for example \(13\) when \(e=1p\):
\begin{equation}
 0=A(\eta)R_{p-1}^{\rm str}+B(\eta).
 \label{app:ear-separation}
\end{equation}
Since \(p-1\ge4\), the independence established above forces
\(A=B=0\), with the other component generic.
All zero-level residues vanish; Section III completes the total-level induction:
\begin{equation}
 R^T_{n,\bmm}=t_nR^{T,\rm str}_{n,\bmm}+q_n,\qquad n\ge5.
 \label{app:two-tower-classification}
\end{equation}

\paragraph{The six-point coefficient example.}
\label{app:six-example}

For the example in the Letter, set \(t_6=1,q_6=0\), as fixed in
Section V. Write \(R=R^{13,14,15}_{1,1,1}\) in independent
variables \((X_{24},X_{35},X_{46};c_{24},c_{35},c_{25})\).
The degree bound \eqref{eq:bootstrap} gives the 13 monomials
\begin{equation}
 \{1,X_{24},X_{35},X_{46},c_{24},c_{35},c_{25},
 X_{24}X_{35},X_{24}X_{46},X_{35}X_{46},
 X_{24}c_{35},X_{46}c_{24},X_{24}X_{35}X_{46}\}.
 \label{app:six-spin}
\end{equation}
The restrictions in Eq.~\eqref{eq:six-boundaries} use
\(c_{13}=X_{24}=0\) and \(c_{46}=X_{46}=0\).
In Fig.~\ref{fig:channels}(b), the first flip uses \(c_{13}=-1\),
\(h_Q=1\). The second has \(h_Q=0\) and
\((k_{25},k_{26},k_{35},k_{36})=(0,0,0,1)\) or \((0,1,0,0)\).
These give \(X_{25}=0\), \(X_{26}=-1\), with \(X_{35}=-1,0\),
respectively. The classification and reference scalar limit
\eqref{app:scalar} fix the endpoint residue to
\((1-X_{35})(1-X_{26})\), as used in the Letter.

\subsection{V. Reconstructing the amplitude}
\refstepcounter{smapp}\label{sm:V}
\label{app:reconstruct}

The Regge condition removes both the constant tower \(q_n\) and
terms invisible to maximal residues. Use the ordered fan
\((13,14,\ldots,1(n-1))\), of size \(D=n-3\), and write
\[
 s_i=X_{1,i+1},\qquad \kappa_i=X_{i,i+2},\qquad 2\le i\le n-2,
 \qquad
 \mathcal L=\{(i,j):2\le i<j\le n-1,\ j-i\ge2\}.
\]
The \(s_i\) are fan channels; \(\kappa_i\) and
\(c_{ij}\), \((i,j)\in\mathcal L\), fix all nonfan channels by
the ABHY identity
\begin{equation}
 X_{ij}=\sum_{\ell=i}^{j-2}\kappa_\ell
 -\sum_{\substack{i\le r<s\le j-1\\s-r\ge2}}c_{rs},
 \qquad 2\le i<j\le n,\quad j-i\ge2.
 \label{app:nonfan-coordinates}
\end{equation}
Thus varying one fan channel holds every other channel fixed.

Let \(f_j\) be the residue in the first \(j\) fan channels at
arbitrary fixed nonnegative levels, with \(f_0=\A_n\).
At the next step put \(p=j+2\) and \(w=s_p\). The spectators
\(\xi\) are the nonfan coordinates and the remaining \(s_i\), \(i>p\).
On their nonempty complex domain
\begin{equation}
 \Omega_p=\{\xi:\ 0<\Re\kappa_i<1,\quad
 \Re c_{ij}<0\ ((i,j)\in\mathcal L),\quad
 \Re s_i>0\ (i>p)\},
 \label{app:regge-domain}
\end{equation}
we require every candidate to obey
\begin{equation}
 \frac1{2\pi}\int_0^{2\pi}
 |f_j(R_Ne^{i\theta},\xi)|\,\dd\theta\longrightarrow0,
 \qquad R_N=N+\tfrac12,\quad 0\le j<D,
 \label{app:prefix-regge}
\end{equation}
locally uniformly in \(\xi\). Constants may depend on the fixed
levels; neither a level-uniform bound nor decay of \(f_D\) is required.
This is stronger than real-axis Regge scaling. Section VI verifies it
for the string solution. The domain is used only for the proof;
joint meromorphy extends the resulting identities.

For \(R_N>|w|\), the finite Cauchy identity is
\begin{equation}
 \begin{aligned}
 f_j(w,\xi)&=\sum_{m=0}^{N}
 \frac{\Res_{z=-m}f_j(z,\xi)}{w+m}+B_{j,N}(w,\xi),\\
 B_{j,N}(w,\xi)&=\frac{1}{2\pi i}
 \oint_{|z|=R_N}\frac{f_j(z,\xi)}{z-w}\,\dd z.
 \end{aligned}
 \label{app:cauchy-descent}
\end{equation}
The boundary integral is bounded by \(R_N/(R_N-|w|)\) times the
average in Eq.~\eqref{app:prefix-regge}, and hence vanishes.
Iterating this finite identity gives the maximal-residue sum and
boundary integrals denoted schematically by \(B_n^T\) in
Eq.~\eqref{eq:pole-reconstruction}; these need not be contact terms.
Take the cutoff limits from the last fan variable to the first.
The earlier level sums then remain finite at each step, so no
level-uniform estimate or interchange of infinite sums is needed.

For \(n\ge5\), take the first \(D-1\) residues at level zero in
\(\A_n-t_n\A_n^{\rm str}\), and call the resulting function of
the last fan variable \(\psi(w,\xi)\).
Equation~\eqref{app:two-tower-classification} makes each residue
of \(\psi\) equal to \(q_n\). The circle encloses the \(N+1\)
possible poles \(0,-1,\ldots,-N\), so
\begin{equation}
 \frac{1}{2\pi iR_N}\oint_{|w|=R_N}\psi(w,\xi)\,\dd w
 =q_n\frac{N+1}{R_N}\longrightarrow q_n.
 \label{eq:constant-tower}
\end{equation}
The left side is bounded in absolute value by the circle average
of \(|\psi|\), which vanishes by Eq.~\eqref{app:prefix-regge}
for the candidate and the string solution. Thus \(q_n=0\),
irrespective of any entire contribution. The all-zero-level string
residue is \(1\), by Eq.~\eqref{app:scalar}; normalization fixes
\(t_n=1\) independently at each multiplicity.

Now take the same residues of \(\A_n-\A_n^{\rm str}\), again
denoting them by \(f_j\). Its maximal residues vanish, so \(f_D=0\).
If \(f_{j+1}=0\) at every next level, all possible pole residues
of \(f_j\) vanish: the other channels stay fixed by
Eq.~\eqref{app:nonfan-coordinates}. The finite Cauchy identity
then leaves only its vanishing boundary integral, giving \(f_j=0\)
on the spectator domain. Descending from \(j=D-1\) to zero proves
equality there, and joint meromorphy extends it to all kinematics.
At four points the same argument applies to the residues in
Eq.~\eqref{eq:beta}.

\subsection{VI. Checking the string solution}
\refstepcounter{smapp}\label{sm:VI}

We check the assumptions and reference identities using the known
disk integral, without imposing this representation on the unknown
amplitude.

\paragraph{Poles, degree bound, and scalar limits.}
\label{app:membership}
The disk integral is
\(\A_n^{\rm str}=\int_{\mathcal M^+_{0,n}}\Omega_{\rm PT}
\prod_\gamma u_\gamma^{X_\gamma}\)
\cite{KobaNielsen,ZDiskIntegrals,NonAbelianZTheory,StringyCanonicalForms,Rigidity}.
Here \(\mathcal M^+_{0,n}\) parametrizes ordered boundary punctures
modulo \(SL(2,\mathbb R)\), \(u_\gamma\) are dihedral cross ratios,
and \(\Omega_{\rm PT}\) contains the Parke--Taylor factor
\(1/[(z_1-z_2)\cdots(z_n-z_1)]\).
Only compatible boundaries meet, with independent normal coordinates;
Mellin continuation gives simple poles at \(X_e=-m\) and no other
finite poles.

At a corner labeled by \(T\), put \(\tau_e=u_e\), \(e\in T\),
and \(\Omega_{\rm PT}=J_T(\tau)\prod_e\dd\tau_e/\tau_e\),
with regular \(J_T\). The physical residue is
\begin{equation}
 R^T_{\boldsymbol m}=[\tau^{\boldsymbol m}]J_T(\tau)
 \exp\!\left(\sum_{\gamma\notin T}X_\gamma\log u_\gamma(\tau)\right).
 \label{app:string-coefficient}
\end{equation}
Each logarithm vanishes at the corner, so the coefficient is polynomial.
The binary identity
\(1-u_\gamma=\prod_{\delta\text{ crosses }\gamma}u_\delta\)
gives \(\log u_\gamma=O(\tau_e)\) when \(\gamma\) crosses \(e\).
More than \(m_e\) derivatives in those crossing variables therefore
annihilate the coefficient, proving the joint bound. At level zero,
the constant term factorizes into the two lower-point polygons:
\(R_3^{\rm str}=1\) and
\begin{equation}
 R^{T,\rm str}_{n,(0,\boldsymbol m_L,\boldsymbol m_R)}
 =R^{T_L,\rm str}_{n_L,\boldsymbol m_L}
  R^{T_R,\rm str}_{n_R,\boldsymbol m_R}.
 \label{app:scalar}
\end{equation}
This factorization is checked only for the reference solution.

\paragraph{Pairing at arbitrary attached levels.}
Take common-channel residues at fixed levels \(h_s\), \(s\in S\),
with other independent kinematics
generic. The four consecutive clusters \(I,J,K,L\) have representatives
at \(t,0,1,\infty\). The endpoints of the physical interval \(t<0\)
are the flipped channels; \(t=1\) is the nonadjacent \(I\)--\(K\)
collision. Use nested internal scales \(\boldsymbol\zeta\), rather
than the dihedral coordinates above, with
\(z_i=Z_A+\delta z_i(\boldsymbol\zeta)\), \(A\in\{I,J,K,L\}\).
Here \(\delta z_i\) is polynomial in scales along the attached tree.
Keep one leaf at each center and use inverted coordinates in the
infinity cluster. Neither gauge fixing nor the Jacobian introduces
a denominator \(1-t\).

On the pairing locus for \(Q=I\times K\), the corresponding
Koba--Nielsen factors become
\begin{equation}
 (z_j-z_i)^{-c_{ij}}=(1-t+\delta z_j-\delta z_i)^{k_{ij}},
 \qquad k_{ij}\in\mathbb Z_{\ge0}.
 \label{app:polynomial-pairs}
\end{equation}
They are polynomial before extraction, so every fixed-level coefficient
is regular at \(t=1\); the Parke--Taylor denominator contains no
\(I\)--\(K\) pair. Other finite differences give powers of \(t\),
and infinity-cluster coefficients are polynomial in the finite centers.
For a cluster \(A\), \(X_A=(\sum_{i\in A}p_i)^2\) is zero for
a singleton or fixed at a common pole. At the paired levels
\(X_e=-m,X_f=-r\), aggregate powers such as \(X_e-X_I-X_J\)
are integers. The remaining one-form \(\omega_S\) is therefore
a finite Laurent sum in \(t\), with poles only at zero and infinity.

Near \(t=0\), put \(x=-t\) and rescale each joining cluster's root
scale as \(\zeta_s=x\eta_s\), leaving descendant relative scales
unchanged. Near infinity, invert the punctures and use \(v=-1/t\).
The representative leaves make \(t\) independent of the internal
scales. For fixed \(t\ne0,\infty\), either change
\(\boldsymbol\zeta=\Phi_t(\boldsymbol\eta)\) is analytic,
invertible, and preserves every common divisor.
At \(X_s=-h_s\), the disk form is
\(\dd t\,g(t,\boldsymbol\zeta)
\prod_s\dd\zeta_s/\zeta_s^{h_s+1}\), with \(g\) holomorphic in
the internal scales. Changing variables on a small product contour
\(\Gamma_\zeta\) around the common divisors gives
\begin{equation}
 \begin{aligned}
 \omega_S(t)
 &=\frac{\dd t}{(2\pi i)^{|S|}}
 \oint_{\Gamma_\zeta}g(t,\boldsymbol\zeta)
       \prod_{s\in S}\frac{\dd\zeta_s}{\zeta_s^{h_s+1}}\\
 &=\frac{\dd t}{(2\pi i)^{|S|}}
 \oint_{\Phi_t^{-1}(\Gamma_\zeta)}
 \frac{g(t,\Phi_t(\boldsymbol\eta))\det D_{\eta}\Phi_t}
      {\prod_{s\in S}\Phi_{t,s}(\boldsymbol\eta)^{h_s+1}}
       \prod_{s\in S}\dd\eta_s .
 \end{aligned}
 \label{app:attached-residue-change}
\end{equation}
The transformed contour is homologous to the product contour around
the same divisors, with its induced orientation. This identifies the
full residue form, including its Jacobian, not a bare Taylor coefficient:
\(\zeta_s=x\eta_s\) contributes \(x^{-h_s}\).
With \(h_A=-X_A\) for a nonsingleton cluster and \(h_A=0\)
otherwise, the joining roots supply
\[
 x^{X_e-X_I-X_J}x^{-h_I-h_J}=x^{X_e}.
\]
Thus the endpoint Mellin factors are \(x^{X_e}\dd x/x\) and,
after inversion, \(v^{X_f}\dd v/v\); their endpoint residues are
the physical maximal residues. Equality on punctured neighborhoods
suffices, without invertibility at the endpoint.
Choosing the first endpoint's physical orientation,
\(\dd x/x=-\dd v/v\) fixes the relative sign:
\begin{equation}
 \Res_{t=0}\omega_S=R^{S\cup\{e\}},\qquad
 \Res_{t=\infty}\omega_S=-R^{S\cup\{f\}}.
 \label{app:endpoint-matching}
\end{equation}
The residue theorem gives pairing; rotating the clusters gives the
other rectangle. This holds at arbitrary common levels, with residues
taken before kinematic restriction. At complete integer resonances,
paired flips and Section II give ADR. Regularity at \(t=1\) was
checked from the string integrand, not inferred from ordinary locality.

\paragraph{Complex Regge decay after fixed-level residues.}
\label{app:regge-proof}
Use the coordinates and domain of Section V. In the gauge
\(z_1=0,z_{n-1}=1,z_n=\infty\), put
\(z_i=\prod_{k=i}^{n-2}t_k\). For \(\Re s_i>0\),
\begin{equation}
 \A_n^{\rm str}=\int_{[0,1]^{n-3}}
 \prod_{i=2}^{n-2}\!\left[\dd t_i\,t_i^{s_i-1}(1-t_i)^{\kappa_i-1}\right]
 \prod_{(i,j)\in\mathcal L}
 \left(1-\prod_{k=i}^{j-1}t_k\right)^{-c_{ij}}.
 \label{app:cube}
\end{equation}
At step \(p\), the fixed preceding residues defining
\(f_{p-2}^{\rm str}\) extract Taylor coefficients in
\(t_2,\ldots,t_{p-1}\) at zero. This inserts a polynomial in
the remaining variables, with no inverse powers of \(1-t_p\):
expanding a factor containing an extracted variable leaves only
nonnegative powers of the others. Localizing at the extracted
boundaries justifies this operation; complementary pieces are
entire in the corresponding Mellin variable. The remaining beta
weights are absolutely and locally uniformly integrable on
\(\Omega_p\), so extraction commutes with their integrals.

Fix a compact spectator set in \(\Omega_p\), put \(t=t_p\),
and let \(\mu=\min\Re\kappa_p\in(0,1)\).
Each factor from the second product that still depends on \(t\)
has the form \((1-tu)^a\), with \(u\in[0,1]\) and uniformly
\(\Re a\ge\delta>0\). Gamma-ratio estimates give
\begin{equation}
 \left|\frac{(-a)_k}{k!}u^k\right|\le C(k+1)^{-1-\delta},
 \qquad
 \left|\frac{(1-\kappa_p)_\ell}{\ell!}\right|
 \le C(\ell+1)^{-\mu}.
 \label{app:coefficient-bound}
\end{equation}
The finitely many small indices are bounded polynomially in the
parameters, including when a series terminates. Finite convolutions
preserve the summable first tail. Convolving it with the second
gives \(O((\nu+1)^{-\mu})\), by splitting at half the total index.
Polynomial insertions shift only finitely many indices, and
integration against the remaining beta weights preserves the bound.
Hence, with \(w=s_p\),
\begin{equation}
 f_{p-2}^{\rm str}(w,\xi)=\sum_{\nu\ge0}\frac{b_\nu(\xi)}{w+\nu},
 \qquad |b_\nu(\xi)|\le C(\nu+1)^{-\mu}.
 \label{app:derived-channel}
\end{equation}
Here and below \(C\) may depend on the fixed preceding levels and
the compact spectator set. For \(\Re w>0\), termwise integration
follows from \(\sum_\nu(\nu+1)^{-\mu}/(\nu+\Re w)<\infty\).
The series then continues meromorphically, locally uniformly away
from its displayed poles.

On \(|w|=R_N=N+\tfrac12\),
\(|w+\nu|\ge|R_N-\nu|\ge\tfrac12\).
Split the sum at \(R_N/2\) and \(2R_N\). The low and high ranges
are \(O(R_N^{-\mu})\); the middle range is bounded by
\(CR_N^{-\mu}\sum_{R_N/2\le\nu\le2R_N}|R_N-\nu|^{-1}
=O(R_N^{-\mu}\log R_N)\). Thus
\begin{equation}
 \sup_{|w|=R_N}|f_{p-2}^{\rm str}(w,\xi)|
 \le C R_N^{-\mu}\log R_N\longrightarrow0.
 \label{app:string-regge}
\end{equation}
This proves Eq.~\eqref{app:prefix-regge} on full complex circles,
uniformly on each compact spectator set, at every fixed tuple of
preceding levels.

%% file: figures/weighted_arms.tex
\begin{tikzpicture}[x=1cm,y=1cm,>=Latex,
 every node/.style={font=\footnotesize},
 line cap=round,line join=round,
 dual edge/.style={draw=resblue!80,densely dashed,line width=.8pt},
 polygon edge/.style={draw=resgray,line width=.9pt},
 schematic edge/.style={draw=resgray!70,line width=.7pt},
 triangle node/.style={circle,fill=resgray,draw=white,line width=.3pt,
   inner sep=0pt,minimum size=4.5pt},
 ear node/.style={triangle node,fill=resblue},
 level/.style={fill=white,inner sep=1pt},
 boundary level/.style={text=resgray,font=\scriptsize},
 vertex label/.style={font=\scriptsize},
 flipped edge/.style={draw=resrust,line width=1.4pt}]
 \path[use as bounding box] (0,-5.70) rectangle (15.40,2.85);

 \begin{scope}[shift={(2.60,.90)}]
  \node at (0,1.73) {\textbf{(a)}\quad Arm lengths $(1,1)$};
  \node[font=\scriptsize] at (0,1.40) {$3$ triangles: a terminal pentagon};
  \coordinate (P1) at (-1.25,-.65);
  \coordinate (P2) at (-1.40,.40);
  \coordinate (P3) at (0,1.10);
  \coordinate (P4) at (1.40,.40);
  \coordinate (P5) at (1.25,-.65);
  \coordinate (outside) at (0,-1.16);
  \coordinate (L) at (-.883,.283);
  \coordinate (C) at (0,-.067);
  \coordinate (R) at (.883,.283);
  \fill[resblue!9] (P1)--(P2)--(P3)--(P4)--(P5)--cycle;
  \fill[resgray!7] (P1)--(P5)--(outside)--cycle;
  \draw[draw=resgray!50,densely dotted,line width=.6pt]
    (P1)--(outside)--(P5);
  \draw[schematic edge] (P1)--(P2)--(P3)--(P4)--(P5)--cycle;
  \draw[schematic edge] (P1)--(P3)--(P5);
  \draw[dual edge] (L)--(C)--(R);
  \draw[dual edge] (C)--(0,-.97);
  \node[level] at (-.925,-.195) {$a$};
  \node[level] at (.925,-.195) {$b$};
  \node[level] at (.73,-.65) {$c$};
  \node[triangle node] at (C) {};
  \foreach \v in {L,R} \node[ear node] at (\v) {};
  \node[text=resgray,font=\scriptsize] at (0,-1.34) {rest of polygon};
 \end{scope}

 \begin{scope}[shift={(7.55,.90)}]
  \node at (0,1.73) {\textbf{(b)}\quad Arm lengths $(1,2)$};
  \node[font=\scriptsize] at (0,1.40) {$4$ triangles};
  \coordinate (P1) at (-1.40,-.75);
  \coordinate (P2) at (-1.55,.40);
  \coordinate (P3) at (-.25,1.10);
  \coordinate (P4) at (1.10,.85);
  \coordinate (P5) at (1.55,-.05);
  \coordinate (P6) at (1.05,-.75);
  \coordinate (outside) at (-.175,-1.16);
  \coordinate (L) at (-1.067,.25);
  \coordinate (C) at (-.20,-.133);
  \coordinate (R) at (.783,.10);
  \coordinate (E) at (.80,.633);
  \fill[resgray!7] (P1)--(P6)--(outside)--cycle;
  \draw[draw=resgray!50,densely dotted,line width=.6pt]
    (P1)--(outside)--(P6);
  \draw[schematic edge] (P1)--(P2)--(P3)--(P4)--(P5)--(P6)--cycle;
  \draw[schematic edge] (P1)--(P3)--(P6) (P3)--(P5);
  \draw[dual edge] (L)--(C)--(R)--(E);
  \draw[dual edge] (C)--(-.183,-1.02);
  \node[level] at (-.997,-.102) {$a$};
  \node[level] at (.634,-.158) {$b$};
  \node[level] at (.49,-.75) {$c$};
  \node[level] at (.344,.721) {$d$};
  \foreach \v in {C,R} \node[triangle node] at (\v) {};
  \foreach \v in {L,E} \node[ear node] at (\v) {};
  \node[text=resgray,font=\scriptsize] at (0,-1.34) {rest of polygon};
 \end{scope}

 \begin{scope}[shift={(12.55,.90)}]
  \node at (0,1.73) {\textbf{(c)}\quad Arm lengths $(2,2)$};
  \node[font=\scriptsize] at (0,1.40) {$5$ triangles};
  \coordinate (P1) at (-1.55,-.75);
  \coordinate (P2) at (-1.60,.08);
  \coordinate (P3) at (-1.10,.82);
  \coordinate (P4) at (0,1.10);
  \coordinate (P5) at (1.10,.82);
  \coordinate (P6) at (1.60,.08);
  \coordinate (P7) at (1.55,-.75);
  \coordinate (outside) at (0,-1.16);
  \coordinate (C) at (0,-.133);
  \coordinate (L) at (-1.05,.143);
  \coordinate (R) at (1.05,.143);
  \coordinate (E) at (-.90,.667);
  \coordinate (F) at (.90,.667);
  \fill[resgray!7] (P1)--(P7)--(outside)--cycle;
  \draw[draw=resgray!50,densely dotted,line width=.6pt]
    (P1)--(outside)--(P7);
  \draw[schematic edge] (P1)--(P2)--(P3)--(P4)--(P5)--(P6)--(P7)--cycle;
  \draw[schematic edge] (P1)--(P4)--(P7) (P2)--(P4)--(P6);
  \draw[dual edge] (E)--(L)--(C)--(R)--(F);
  \draw[dual edge] (C)--(0,-1.02);
  \node[level] at (-.72,.241) {$a$};
  \node[level] at (.72,.241) {$b$};
  \node[level] at (.78,-.75) {$c$};
  \node[level] at (-.64,.692) {$d$};
  \node[level] at (.64,.692) {$e$};
  \foreach \v in {C,L,R} \node[triangle node] at (\v) {};
  \foreach \v in {E,F} \node[ear node] at (\v) {};
  \node[text=resgray,font=\scriptsize] at (0,-1.34) {rest of polygon};
 \end{scope}

 \node at (7.70,-1.09)
   {\textbf{(d)}\quad The $(1,2)$ move for $b\ge a$: triangulation and dual tree};
 \node at (3.15,-1.53) {Before: four triangles};
 \node at (12.05,-1.53) {After: a terminal pentagon};

 \foreach \panel/\xc in {before/3.15,after/12.05} {
  \begin{scope}[shift={(\xc,-3.35)}]
   \coordinate (P1) at (-1.62,-.63);
   \coordinate (P2) at (-1.62,.63);
   \coordinate (P3) at (0,1.25);
   \coordinate (P4) at (1.62,.63);
   \coordinate (P5) at (1.62,-.63);
   \coordinate (P6) at (0,-1.25);
   \coordinate (outside) at (-1.43,-1.57);
   \coordinate (earA) at (-1.08,.417);
   \coordinate (earD) at (1.08,.417);
   \def\afterpanel{after}
   \ifx\panel\afterpanel
    \fill[resblue!9] (P1)--(P2)--(P3)--(P4)--(P5)--cycle;
    \coordinate (innerA) at (0,-.003);
    \coordinate (innerD) at (0,-.837);
   \else
    \coordinate (innerA) at (-.54,-.21);
    \coordinate (innerD) at (.54,-.21);
   \fi
   \fill[resgray!7] (P1)--(P6)--(outside)--cycle;
   \draw[draw=resgray!50,densely dotted,line width=.6pt]
     (P1)--(outside)--(P6);
   \draw[polygon edge] (P1)--(P2)--(P3)--(P4)--(P5)--(P6)--cycle;
   \draw[polygon edge] (P1)--(P3)--(P5);
   \ifx\panel\afterpanel
    \draw[flipped edge] (P1)--(P5);
   \else
    \draw[flipped edge] (P3)--(P6);
   \fi
   \draw[dual edge] (earA)--(innerA);
   \ifx\panel\afterpanel
    \draw[dual edge] (earD)--(innerA);
   \else
    \draw[dual edge] (earD)--(innerD);
   \fi
   \draw[dual edge] (innerA)--(innerD);
   \ifx\panel\afterpanel
    \draw[dual edge] (innerD)--(-1.017,-1.15);
   \else
    \draw[dual edge] (innerA)--(-1.017,-1.15);
   \fi
   \foreach \v in {innerA,innerD} \node[triangle node] at (\v) {};
   \foreach \v in {earA,earD} \node[ear node] at (\v) {};
   \node[level] at (-1.03,-.10) {$a$};
   \node[level] at (1.03,-.10) {$d$};
   \node[level] at (-1.13,-.81) {$c$};
   \ifx\panel\afterpanel
    \node[level,text=resrust] at (.63,-.63) {$b$};
   \else
    \node[level,text=resrust] at (0,-.65) {$b$};
   \fi
   \node[boundary level] at (-1.80,.03) {$0$};
   \node[boundary level] at (-.84,1.10) {$0$};
   \node[boundary level] at (.84,1.10) {$0$};
   \node[boundary level] at (1.80,.03) {$0$};
   \node[boundary level] at (.91,-1.06) {$0$};
   \node[vertex label,anchor=east] at (-1.70,-.65) {$1$};
   \node[vertex label,anchor=east] at (-1.70,.67) {$2$};
   \node[vertex label,anchor=south] at (0,1.31) {$3$};
   \node[vertex label,anchor=west] at (1.70,.67) {$4$};
   \node[vertex label,anchor=west] at (1.70,-.65) {$5$};
   \node[vertex label,anchor=north] at (0,-1.31) {$6$};
   \node[text=resgray,font=\scriptsize] at (-1.48,-1.78) {rest of polygon};
  \end{scope}
 }

 \draw[->,resgray,line width=.9pt] (5.60,-3.35)--(9.55,-3.35);
 \node at (7.57,-2.93) {flip $36\longrightarrow15$, same level $b$};
 \node at (7.57,-3.77) {$\min(a,c+d)\le b$};
 \node[align=center,text=resblue] at (7.57,-4.57)
   {$3$ triangles $=$ a pentagon\\$4$ original sides $+$ attachment $b$};

 \draw[polygon edge] (1.70,-5.54)--(2.22,-5.54);
 \node[anchor=west,font=\scriptsize] at (2.30,-5.54) {triangulation};
 \draw[dual edge] (5.03,-5.54)--(5.60,-5.54);
 \node[triangle node] at (5.03,-5.54) {};
 \node[triangle node] at (5.60,-5.54) {};
 \node[anchor=west,font=\scriptsize] at (5.69,-5.54) {dual tree: one dot per triangle};
 \node[ear node] at (11.29,-5.54) {};
 \node[anchor=west,font=\scriptsize] at (11.45,-5.54) {ear (two original sides)};
\end{tikzpicture}